# Evolution of major sedimentary mounds on Mars:
# build-up via anticompensational stacking modulated by climate change

Edwin S. Kite[1,*], Jonathan Sneed[1], David P. Mayer[1], Kevin W. Lewis[2], Timothy I. Michaels[3], Alicia Hore[4], Scot C.R. Rafkin[5].

1. University of Chicago. 2. Johns Hopkins University. 3. SETI Institute. 4. Brock University.
5. Southwest Research Institute. (*kite@uchicago.edu)

## Abstract.

We present a new database of >300 layer-orientations from sedimentary mounds on Mars. These layer orientations, together with draped landslides, and draping of rocks over differentially-eroded paleo-domes, indicate that for the stratigraphically-uppermost ~1 km, the mounds formed by the accretion of draping strata in a mound-shape. The layer-orientation data further suggest that layers lower down in the stratigraphy also formed by the accretion of draping strata in a mound-shape. The data are consistent with terrain-influenced wind erosion, but inconsistent with tilting by flexure, differential compaction over basement, or viscoelastic rebound. We use a simple landscape evolution model to show how the erosion and deposition of mound strata can be modulated by shifts in obliquity. The model is driven by multi-Gyr calculations of Mars' chaotic obliquity and a parameterization of terrain-influenced wind erosion that is derived from mesoscale modeling. Our results suggest that mound-spanning unconformities with kilometers of relief emerge as the result of chaotic obliquity shifts. Our results support the interpretation that Mars' rocks record intermittent liquid-water runoff during a $\gg 10^8$-yr interval of sedimentary rock emplacement.

## 1. Introduction.

Understanding how sediment accumulated is central to interpreting the Earth's geologic records (Allen & Allen 2013, Miall 2010). Mars is the only other planet known to host an extensive sedimentary record. Gale crater and the Valles Marineris (VM) canyon system contain some of Mars' thickest (2-8km) and best-exposed sequences of sedimentary rock (Malin & Edgett 2000, Milliken et al. 2010). "The origin of [these sedimentary] mounds is a major unresolved question in Mars geology" (Grotzinger & Milliken 2012). The mounds are thought to have formed <3.7 Ga, relatively late in Mars' aqueous history and many contain sulfates that precipitated from aqueous fluids (Gendrin et al. 2005, Bibring et al. 2006, Mangold et al. 2008, Murchie et al. 2009a). The fluid source could be groundwater, rain, or snowmelt (Andrews-Hanna et al. 2010, Kite et al. 2013b). Proposed depositional scenarios (Nedell et al. 1987, Lucchita 1992) range from primarily aeolian sedimentation in a climate dry enough that aeolian erosion could define moats around the growing mounds (Catling et al. 2006, Michalski & Niles 2012, Kite et al. 2013a), through sand/dust cementation in horizontal playa-lake beds (Andrews-Hanna et al. 2010, Fueten et al. 2008, Murchie et al. 2009b), to fluvial sediment transport from canyon/crater rims into canyon/crater-spanning lakes (Grotzinger et al. 2015). At Gale crater, aeolian processes contributed to the deposition of the mound, evidenced by preserved bedforms within the stratigraphy (Milliken et al. 2014, Banham et al. 2016). Following the depositional era, aeolian erosion cut into the rocks, exposing layers and perhaps deepening moats (Day & Kocurek 2016).

These paleo-environmental scenarios make contrasting predictions for the orientations of mound sediment layers and unconformities. If layers dip away from mound crests and layers have been little tilted since the time of deposition, then the mounds formed as mounds (similar to ice mounds within Mars' polar craters; e.g. Brothers & Holt 2016). By contrast, gravity-driven deposition predicts layers that were originally flat-lying or oriented away from crater walls or canyon walls, with the modern topography resulting entirely from later erosion. The full internal architecture of the VM and Gale mounds cannot be directly observed, but can be inferred from outcrop measurements of layer-orientations and unconformities (Okubo et al. 2008).

Layer-orientation measurements for Mars are obtained using orbiter image stereopairs to construct digital terrain models (DTMs) that form the basis for fitting planes to traces of stratigraphic surfaces (Lewis et al. 2008). From orbit it is usually not possible (due to limited resolution) to distinguish the traces of beds from the traces of lower-order bounding surfaces, although both should closely correspond to basin topography at around the time of deposition. These fitted planes usually dip in a downslope direction, but may suffer from downslope bias (e.g. Fueten et al. 2006). However, consensus on the interpretation of Mars layer data has been hindered by doubts about the accuracy of layer-orientations measured from orbiter image data, the possibility that layer-orientations do not reflect paleo slopes, and the absence of a physical mechanism that could account both for layer-orientations and for Mars' large unconformities.

Corresponding to this lack of consensus in data interpretation, there are two endmember views of how Mars' mounds formed:

1. In one view, craters/canyons were fully filled by flat-lying or shallowly dipping strata, e.g. playa-deposits or fluviodeltaic deposits (Fig. 1a), and later underwent extensive erosion to their present form (Malin & Edgett 2000, Andrews-Hanna et al. 2010), presumably through wind erosion (Kite et al. 2013a, Day et al. 2016). In this view, the primary cause of non-horizontal layer-orientations is downslope measurement bias and/or postdepositional distortion (by flexure, landslides, soft-sediment deformation, tectonics, and differential compaction) (e.g. Nedell et al. 1987, Metz et al. 2010, Grotzinger et al. 2015). Preferential infilling of topographic lows through deposition (compensational stacking) is ubiquitous in well-studied aqueous sedimentary environments on Earth (Straub et al. 2009). Therefore, it is tempting to assume that Earth analogy, which has been used effectively to interpret sedimentary structures viewed by rovers (McLennan & Grotzinger 2008, Grotzinger et al. 2015), also holds at the scale of Mars basins.

2. In another view, the downslope layer-tilts are primary. If this is correct, then mounds grew in place by net deposition of layers on preexisting topographic highs (anticompensational stacking) (e.g. Niles & Michalski 2012, Kite et al. 2013a). This distinctively Martian mechanism is suggested by:- (i) growth of polar ice/dust/sand mounds by anticompensational stacking (Holt et al. 2010, Conway et al. 2012, Brothers et al. 2013, Brothers & Holt 2016); (ii) the importance of aeolian sediment transport and slope-winds on modern Mars (Spiga et al. 2011, Spiga 2011, Kok et al. 2012, Bridges et al. 2013, Silvestro et al. 2013, Kite et al. 2013a); (iii) the strong inference of layered-sediment accumulation via anticompensational stacking for some Mars equatorial layered sediments (the Medusae Fossae Formation; Bradley et al. 2002, Zimbelman & Scheidt 2012, Kite et al. 2015). Dry conditions bring aeolian processes to the fore, whereas vigorous and sustained fluvial erosion would inhibit mound construction. Therefore anticompensational stacking corresponds to a paleoenvironment where fluvial sediment transport is infrequent, consistent with models of Mars paleoclimate (Kite et al. 2013b, Mischna et al. 2013, Segura et al. 2013, Urata and Toon 2013, Halevy and Head 2014,

Wordsworth et al. 2013, Ramirez et al. 2014, Kerber et al. 2015, Wordsworth et al. 2015, Wordsworth 2016).

## 1.1. Outline.

Here we construct a new database (section 2) of layer-orientations (section 3) and unconformities (section 4) within Martian mounds, in order to constrain accumulation of sedimentary rocks (section 5). We also present a new model (section 6) of mound emplacement. Implications and tests are discussed in section 7, and conclusions are listed in section 8.

Our work has 3 purposes:

a) *To address concerns to the mounds-grew-as-mounds hypothesis of Kite et al. (2013a).* These concerns are:-

- That layer orientations "have not been independently confirmed" (Grotzinger et al. 2015);
- That layer orientations can be accounted for by differential compaction of originally-horizontal layers over basement relief (basement = rocks that predate sedimentary infill), removing the need for slope- wind erosion during the depositional era (Grotzinger et al. 2015).

We resolve these concerns in sections 2-5:-

- Exhaustive tests show that layer orientations are accurate and reproducible, and that layer orientations errors (including downslope bias) are insignificant for the purpose of determining mound origin (section 2).
- Layer-orientations in VM mounds show an outward dip – a direction opposite that predicted for differential compaction over basement relief (section 3). Layer-orientations in Gale are unlikely to result from differential compaction over a central ring or central peak (Gabasova & Kite 2016) (section 3). Unconformity data and draped-landslide data show that the mounds grew by anticompensational stacking at least for the topmost ~1 km of the mounds (section 4). Below this level, the data suggest two options: (i) accretion of draping strata in a mound shape, or (ii) slope-wind erosion sculpts pre-compacted sedimentary deposits, which subsequently act as a mound-shaped form over which later sediments may be differentially compacted (section 5).

b) *To expand the database of High Resolution Imaging Science Experiment (HiRISE)-derived layer orientation data for Mars.* Using HiRISE (McEwen et al. 2007) data, we gathered 182 new layer-orientations from seven VM mounds, and increased the number of independent layer orientations for the mound in Gale crater from 80 to 126, for a total of 308 (section 3). Together, these mounds make up ~½ of the total volume of canyon/crater-hosted sedimentary mounds on Mars. Our work builds on previous studies (e.g. Fueten et al. 2006), but uses a procedure that is more accurate, includes error bars, and has been validated (section 2). Our database can be applied to many Mars geology problems (Supplementary Table). As one example, we test the prediction of Kite et al. (2013a) that systematically outward-oriented dips should be common in Mars mounds (section 5).

c) *To propose a new model for the major unconformities in Mars' mounds.* Our new analysis of stratigraphic surfaces previously-reported as mound-spanning unconformities show that these commonly have a dome shape (section 4). In order to match these data, we introduce a new model (section 6) that quantitatively integrates temporal variations and spatial variations in Mars sedimentation - for the first time for ancient Mars sedimentary-mound analysis (see also Howard 2007). Our model successfully reproduces the shape and tilt of the observed unconformities (section 6).

To support these goals, we improved the `SWEET` (Slope-Wind Enhanced Erosion and Transport) model of Kite et al. (2013a). Kite et al. (2013a) showed how slope-winds create mounds, provided that the crater/canyon is larger than a critical size, and that long-term-average deposition rate is neither much larger nor much smaller than long-term-average wind-erosion rate (consistent with data: Bridges et al. 2012, Lewis & Aharonson 2014). Two features inherent to the relatively simple `SWEET` model are the absence of a physically realistic relationship between slope and shear stress, together with the lack of any explanation of mound-spanning unconformities. (Although steady forcing in `SWEET` can produce autogenic unconformities, the younger layers grow off to one side – rather than building on top of the thickest point of the main mound, as is commonly observed for Mars' mountains.) We solve this problem in our improved model, which we term `SOURED` (Stratigraphy with Obliquity-triggered Unconformities and Relief-influenced Erosion & Deposition). Specifically, we include realistic multi-Gyr calculations of Mars obliquity, and a more-realistic parameterization of terrain-influenced wind erosion derived from mesoscale modeling (Appendix A and Appendix B).

## 1.2. Geologic scope and geologic context.

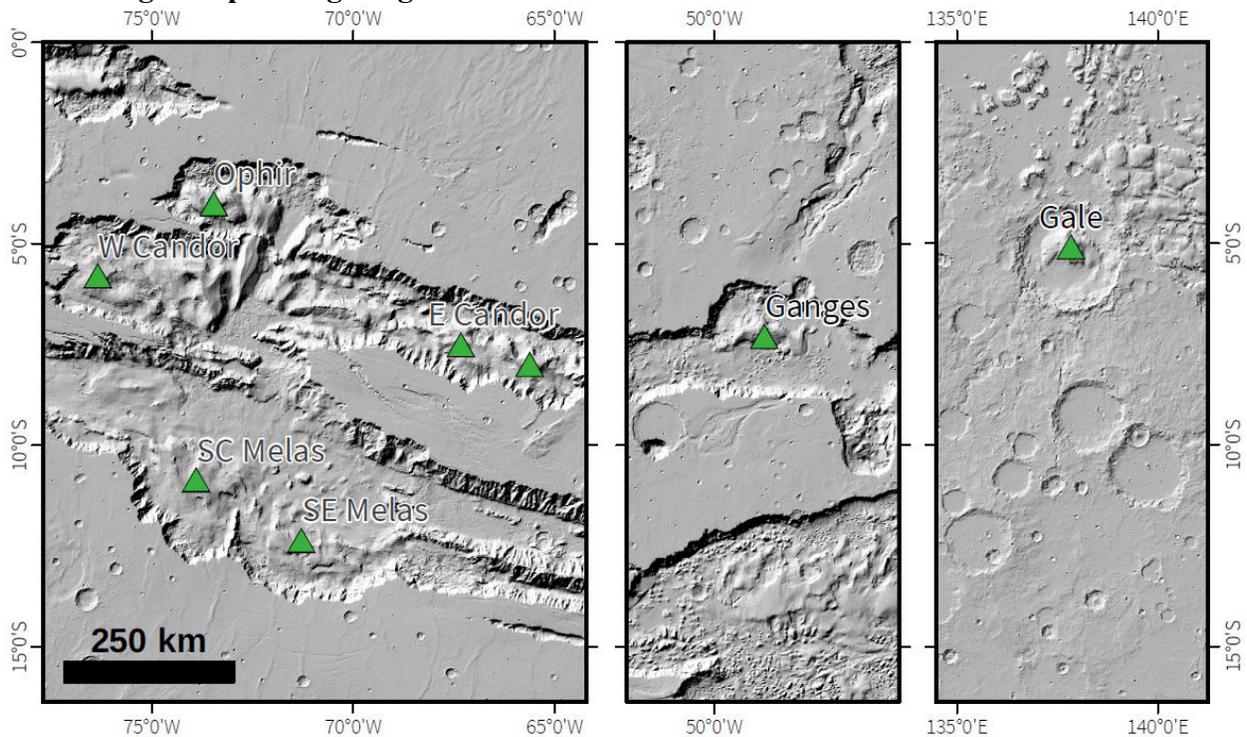

**Fig. 1.** Location of mounds (green triangles) investigated in this work. Background is Mars Orbiter Laser Altimeter shaded relief.

For this study we selected sedimentary mounds that are voluminous, light-toned, show well-exposed off-horizontal layering, have good HiRISE stereopair coverage, and either host sulfates or are stratigraphically associated with sulfates. We further selected only mounds that sit within deep, wide and steep-sided craters/canyons, attributes that favor slope winds. In both VM and Gale, the erodible sedimentary mounds are contained within craters/canyon walls made up of much-less-erodible basement materials. The 8 mounds (mensae) that were selected are Nia (7.6°S, 67.2°W), Juventae (8.0°S, 65.6°W), Mt. Sharp / Aeolis Mons (5.1°S, 137.8°E), Ophir (4.0°S, 73.5°W), Ceti (6.1°S, 75.8°W), Melas (10.7°S, 74.1°W), Coprates (12.5°S, 71.5°W), and Ganges (7.2°S, 48.9°W) – all at <15° latitude. We excluded the well-studied mounds Juventae Chasma (Catling et al. 2006, Bishop et al. 2009) and Candor Mensa (Mangold et al. 2008, Ferguson et al. 2015, Fueten et al. 2014). (The criteria exclude a large number of sedimentary accumulations on Mars: e.g., plateau deposits (Mawrth, Loizeau et al. 2015; Meridiani Planum, Hynek & Phillips 2008), the clay-bearing Terby deposits (Ansan et al. 2011), the free-standing Medusae Fossae mounds (Bradley et al. 2002, Zimbelman & Scheidt 2012, Kite et al. 2015), and veneers and smaller mounds in and around VM (e.g. Milliken et al. 2008, Thollot et al. 2012, Weitz & Bishop 2016).) Gale's mound (Aeolis Mons; also known as Mount Sharp) is the largest among 50 documented crater-hosted mounds outside the polar regions (Bennett & Bell 2016), is the primary science target of MSL (MSL Extended Mission Plan 2014), and has the best HiRISE stereopair coverage of any within-crater mound, justifying our emphasis on this within-crater mound.

The 8 mounds studied here were among the first Mars sedimentary rock accumulations to be described (Malin & Edgett 2000, Malin et al. 2010). The rocks formed relatively late in Mars' aqueous history and contain hematite and sulfates (Christensen et al. 2001, Gendrin et al. 2005, Bibring et al. 2007, Weitz et al. 2008, Murchie et al. 2009a, Roach et al. 2010, Fassett & Head 2011, Ehlmann et al. 2011, Fergason et al. 2014). Most of our layer-orientation data comes from the VM mounds ("Interior Layered Deposits"; ILD). Crosscutting relationships, and contrasts in texture, thermal inertia, erodibility, and mineralogy between sedimentary-mound rocks and canyon-wall rocks all indicate that the ILD accumulated after the canyons formed (Peterson 1981, Lucchitta 2010, Okubo et al. 2008, Schultz 2002, Andrews-Hanna 2012a; see also Montgomery et al. 2009).

Mound stratigraphy (Fig. 2), which usually includes at least one mound-spanning unconformity, is described in a large literature (e.g. Malin & Edgett 2000, Le Deit et al. 2013, Anderson & Bell 2010, Milliken et al. 2010, Thomson et al. 2011, Grotzinger & Milliken 2012). The following trends are a useful guide to correlation: (1) Within-mound materials below the lowest mound-spanning unconformity usually, but not exclusively, correspond to the "Laterally Continuous Sulfate" orbital facies of Grotzinger & Milliken (2012). (2) Materials found above the lowest mound-spanning unconformity within a mound usually, but not exclusively, correspond to the "rhythmite" of Grotzinger & Milliken 2012. (3) Darker-toned indurated materials draping the present topography usually correspond to the widespread "thin mesa" units of Malin & Edgett (2000) (Fig. 2). We did not measure layer-orientations on "thin mesa" units.

The VM and Gale mounds are no older than the Noachian/Hesperian boundary, based on the timing of VM formation, and on the crater-retention age of Gale's ejecta (Anderson et al. 2001, Thomson et al. 2011, Le Deit et al. 2013). The topmost sedimentary rocks could be as young as Upper Amazonian (Mangold et al. 2010, Thollot et al. 2012). Therefore, crater chronology permits a ≫100 Myr interval of sedimentary rock accumulation. This is consistent with other methods (Lewis & Aharonson 2014).

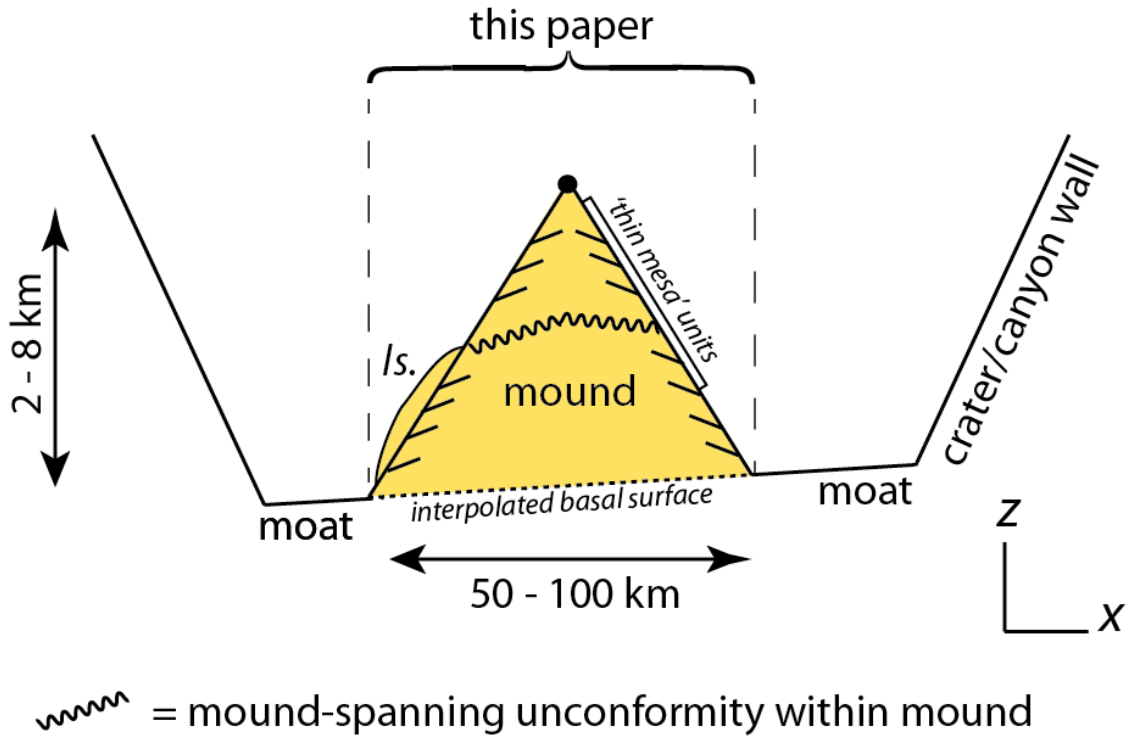

**Fig. 2.** Schematic shows an idealized sedimentary rock mound within an erosion-resistant container (crater or canyon). In this paper, we focus on rocks within the topographically-defined mound (ignoring moat rocks and wall rocks). We neglect <<100m thick "thin mesa" units that drape modern topography (white outline). Layer orientations constrain mode(s) of mound emplacement (section 3). Unconformity data and associated isochores, as well as draped landslides ("ls."), further constrain basin evolution for the upper part of the mounds (section 4).

The physical processes and patterns of deposition for all these rock units is uncertain, and in the words of Grotzinger & Milliken (2012), "[m]easurements of the strike, dip, and stratal geometries of layers within these units would help to place further constraints on their mode(s) of emplacement." Such measurements are the focus of the work presented here. Remarkably, despite the size of the mound-spanning unconformities of Mars, we are not aware of any previous physical model for their origin.

### 1.3. Relation to rover data.

Co-analysis of rover data and orbiter data can increase the science value of both (Arvidson et al. 2006, Fraeman et al. 2013, Arvidson et al. 2015, Lapotre et al. 2016, Stack et al. 2016). In 2012 the Mars Science Laboratory (MSL) rover landed successfully in Gale crater, ~6 km away from the layers in Mt. Sharp / Aeolis Mons where layer-orientations are reported (Le Deit et al. 2013, Kite et al. 2013a, Stack et al. 2013). Rover results to date from Gale crater are interpreted as primarily fluviolacustrine deposits (Grotzinger et al. 2014), which are overlain unconformably by later aeolian sands (Lewis et al. 2015). Our unconformity results based on analysis of orbiter data echo recent rover discoveries in the Gale moat (Watkins et al. 2016). As MSL continues its drive (MSL Extended Mission Plan 2014), the rover's instruments may decisively constrain the sediment transport mechanism for the lower layers of Gale crater's mound (section 7). As of mid-2016, the rover is ~5 km from the sulfate-bearing layers where layer-orientations are reported. Throughout the traverse to date, Mastcam rover imagery has resolution at the sulfate-bearing layers where layer-orientations are reported that is inferior to HiRISE. Specifically, the Mastcam M100 has an angular resolution of 74 µrad/pixel (85 cm/pixel for a 25° slope, 37 cm/pixel for a vertical target at 5 km) (Malin et al. 2010). This compares to HiRISE (from 250 km: 28 cm/px for a 25° slope, 25 cm/px for a horizontal target). Due to foreground obstructions, and edge-on views, layers are more easily visualized in orbiter imagery. The ChemCam Remote Micro-Imager (RMI) has a nominal resolution of 20 µrad/pixel and its potential for long-range stereophotogrammetry is exciting (Le Mouelic et al. 2015). However, we are not aware of any suitable Mt. Sharp / Aeolis Mons RMI stereopairs. Because of the (current) superiority of orbiter images compared to rover images for the purposes of stereo determination of layer-orientations within the sulfate-bearing layers, our work is largely based on orbiter data analysis. Rover imagery shows apparent dips in Mt. Sharp / Aeolis Mons layers that are qualitatively consistent with dips obtained from HiRISE DTMs.

## 2. Data analysis methods.

### 2.1. DTM production method.

HiRISE DTMs and orthoimages were used as the basis for layer tracing (section 1.3). We produced CTX and HiRISE DTMs using the NASA Ames Stereo Pipeline (ASP) (Moratto et al. 2010, Beyer et al. 2014, Shean et al. 2016). As part of this processing, we developed a set of scripts that act as wrappers around the ASP routines, which increase the level of automation and computational efficiency of the DTM production (Mayer & Kite 2016). Initial CTX point clouds were aligned to MOLA shot data using an iterative closest points (ICP) algorithm before being interpolated to DTMs and orthoimages with a grid spacing of 18 m. Initial HiRISE point clouds were then similarly aligned to the CTX DTMs before being interpolated to DTMs and orthoimages with a grid spacing of 1 m (2 m for HiRISE input collected in 2×2 binning mode; Table 1).

As an independent check on the quality of our DTM production workflow, we compared 3 of our HiRISE DTMs to DTMs generated from the same HiRISE stereopairs and available from the Planetary Data System (these PDS DTMs were produced using SOCET SET; Kirk et al. 2008). Because we are primarily interested in the vertical differences between DTMs produced using different methods, we coregistered the PDS-released products to our products by using tie points selected manually on the orthoimages and then applying the resulting transform to the DTMs in order to eliminate any horizontal offsets. We then subtracted the elevation values of our DTMs from the PDS-released DTMs to create a series of difference rasters. For the purposes of the layer orientation measurements in this paper, the most important differences were broad tilts across the entire image. We inspected the resulting difference rasters to characterize tilts. These tilts were 0.2°, 0.17°, and 0.09° for the 3 DTMs investigated, which is much smaller than our error bars.

In addition to the stereo DTMs, digital models of each mound were extracted from MOLA gridded data. Mound basal surfaces were defined from the MOLA elevation data using cubic polynomial interpolation within mound edges. Mound crest-lines and edges were drawn by visual inspection of THEMIS mosaics.

## 2.2. Layer tracing method.

Layer traces were carried out by visual inspection using orthorectified HiRISE images and corresponding DTMs (Fig. 3), following the method of Lewis et al. (2008). Most traces were >150m long. For Gale's mound, we included data from Kite et al. (2013a). Layer orientations were calculated for the best-fit plane for each layer trace. Layers were rejected if their pole error was >2° (calculated following Lewis et al. 2008).

| Crater/ canyon | Mound | DTM location | Image 1 | Image 2 | DTM Posting |
|---|---|---|---|---|---|
| *DTMs produced for this study* | | | | | |
| SE Melas | Coprates Mensa | 13S 289E | ESP_027723_1670 | ESP_027746_1670 | 1 m/pixel |
| SE Melas | Coprates Mensa | 13S 290E | ESP_035450_1670 | ESP_034250_1670 | 1 m/pixel |
| SE Melas | Coprates Mensa | 13S 288E | ESP_028567_1680 | ESP_027657_1680 | 2 m/pixel |
| Ophir | Ophir Mensa | 4S 286E | ESP_034949_1760 | ESP_034738_1760 | 1 m/pixel |
| Ophir | Ophir Mensa | 4S 286E | ESP_015974_1760 | ESP_020220_1760 | 1 m/pixel |
| Ophir | Ophir Mensa | 4S 286E | PSP_008893_1760 | PSP_008458_1760 | 1 m/pixel |
| Ophir | Ophir Mensa | 4S 286E | ESP_017886_1760 | ESP_017675_1760 | 1 m/pixel |
| SC Melas | Melas Mensa | 10S 286E | PSP_010660_1700 | PSP_007812_1700 | 1 m/pixel |
| SC Melas | Melas Mensa | 10S 286E | PSP_005953_1695 | PSP_002630_1695 | 1 m/pixel |
| SC Melas | Melas Mensa | 11S 286E | PSP_001377_1685 | PSP_001852_1685 | 2 m/pixel |
| SC Melas | Melas Mensa | 11S 286E | ESP_012361_1685 | ESP_012572_1685 | 2 m/pixel |
| SC Melas | Melas Mensa | 11S 285E | ESP_033169_1690 | ESP_032747_1690 | 1 m/pixel |
| SC Melas | Melas Mensa | 10S 285E | ESP_028633_1695 | ESP_034382_1695 | 1 m/pixel |
| W Candor | Ceti Mensa | 6S 283E | PSP_003896_1740 | PSP_002841_1740 | 1 m/pixel |
| E Candor | Juventae Mensa | 8S 294E | ESP_017411_1715 | ESP_017266_1715 | 1 m/pixel |
| E Candor | Juventae Mensa | 7S 294E | ESP_037586_1725 | ESP_037731_1725 | 1 m/pixel |
| E Candor | Nia Mensa | 8S 293E | ESP_034896_1725 | ESP_036452_1725 | 1 m/pixel |
| E Candor | Nia Mensa | 7S 292E | ESP_031982_1730 | ESP_031916_1730 | 1 m/pixel |
| E Candor | Nia Mensa | 7S 292E | ESP_014154_1730 | ESP_014431_1730 | 2 m/pixel |
| Gale | Mt. Sharp / Aeolis Mons | 5S 137E | PSP_006855_1750 | PSP_007501_1750 | 1 m/pixel |
| Gale | Mt. Sharp / Aeolis Mons | 5S 137E | ESP_012195_1750 | ESP_012340_1750 | 1 m/pixel |
| Gale | Mt. Sharp / Aeolis Mons | 6S 138E | PSP_003176_1745 | PSP_002464_1745 | 1 m/pixel |
| Gale | Mt. Sharp / Aeolis Mons | 5S 138E | ESP_016375_1750 | ESP_016520_1750 | 1 m/pixel |
| Gale | Mt. Sharp / Aeolis Mons | 5S 138E | ESP_030880_1750 | ESP_030102_1750 | 1 m/pixel |
| Gale | Mt. Sharp / Aeolis Mons | 5S 137E | ESP_012907_1745 | ESP_013540_1745 | 1 m/pixel |
| *Additional DTMs from Kite et al. (2013a), produced and analyzed by K.W. Lewis. 2° (worst-case) error assumed.* | | | | | |
| Gale | Mt. Sharp / Aeolis Mons | 5S 138E | PSP_008437_1750 | /PSP_008938_1750 | 1 m/pixel |
| Gale | Mt. Sharp / Aeolis Mons | 5S 137E | ESP_023957_1755 | ESP_024023_1755 | 1 m/pixel |
| Gale | Mt. Sharp / Aeolis Mons | 5S 137E | PSP_001488_1750 | PSP_001752_1750 | 1 m/pixel |
| Gale | Mt. Sharp / Aeolis Mons | 5S 137E | PSP_009149_1750 | PSP_009294_1750 | 1 m/pixel |
| Gale | Mt. Sharp / Aeolis Mons | 6S 138E | ESP_014186_1745 | ESP_020410_1745 | 1 m/pixel |
| *Additional DTM produced and traced by Okubo (2014), traces not included in the main database.* | | | | | |
| W Candor | Ceti Mensa | 7S 284E | PSP_001641_1735 | PSP_002063_1735 | 1 m/pixel |

| *Additional DTMs produced and analyzed by Alicia Hore (Hore 2015), summarized in Fig. 8f but not included in the main database* | | | | | |
|---|---|---|---|---|---|
| Ganges | Ganges Mensa | 7S 311E | PSP_006519_1730 | PSP_007020_1730 | *n.a.* |
| Ganges | Ganges Mensa | 7S 311E | ESP_013059_1725 | ESP_012993_1725 | *n.a.* |
| Ganges | Ganges Mensa | 7S 311E | PSP_002550_1725 | PSP_003618_1725 | *n.a.* |
| Ganges | Ganges Mensa | 7S 312E | ESP_011648_1730 | ESP_011582_1730 | *n.a.* |
| Ganges | Ganges Mensa | 7S 311E | ESP_018162_1730 | ESP_018633_1730 | *n.a.* |
| Ganges | Ganges Mensa | 7S 311E | PSP_007877_1725 | PSP_007521_1725 | *n.a.* |

**Table 1.** Table of DTMs.

Layers were traced on the HiRISE DTMs listed in Table 1. Linear subhorizontal features observed in Mars outcrops from orbit might correspond to depositional beds, first-order bounding surfaces, deflation surfaces, diagenetic bands, or even buttress unconformities or wave runup features (Rubin & Hunter 1982, Kocurek 1988, Edgar et al. 2012, Parker et al. 2014). Where rovers have explored sulfate-rich rocks on Mars, shallow/early diagenesis blurs the distinction between diagenetic bands and depositional beds. (Later diagenetic fronts need not be parallel to depositional beds; Davies & Cartwright 2002, Borlina et al. 2015). Therefore, we aimed to trace stratigraphic surfaces that closely corresponded to basin-scale topography at the time of deposition (we refer to these stratigraphic surfaces as "layers"). To maximize the likelihood of tracing layers, we followed Lewis (2009) and avoided drawing traces that crossed faults in the rocks where displacement may have occurred, and areas adjacent to faults where folding can distort layers into non-planar surfaces. We avoided tracing on landslides, convolute folding (Metz et al. 2010), superscoops, zones of apparent soft-sediment deformation, and 'thin mesa' materials (Malin & Edgett 2000). Examples of the trace locations and corresponding results are shown in Fig. 3. The fine scale and high degree of lateral continuity of layers (e.g. Fig. 3) is strong evidence that the observed layering represents true depositional bedding and not, for instance, diachronous facies boundaries or late-diagenetic alteration horizons (Le Deit et al. 2013, Stack et al. 2013, Milliken et al. 2014).

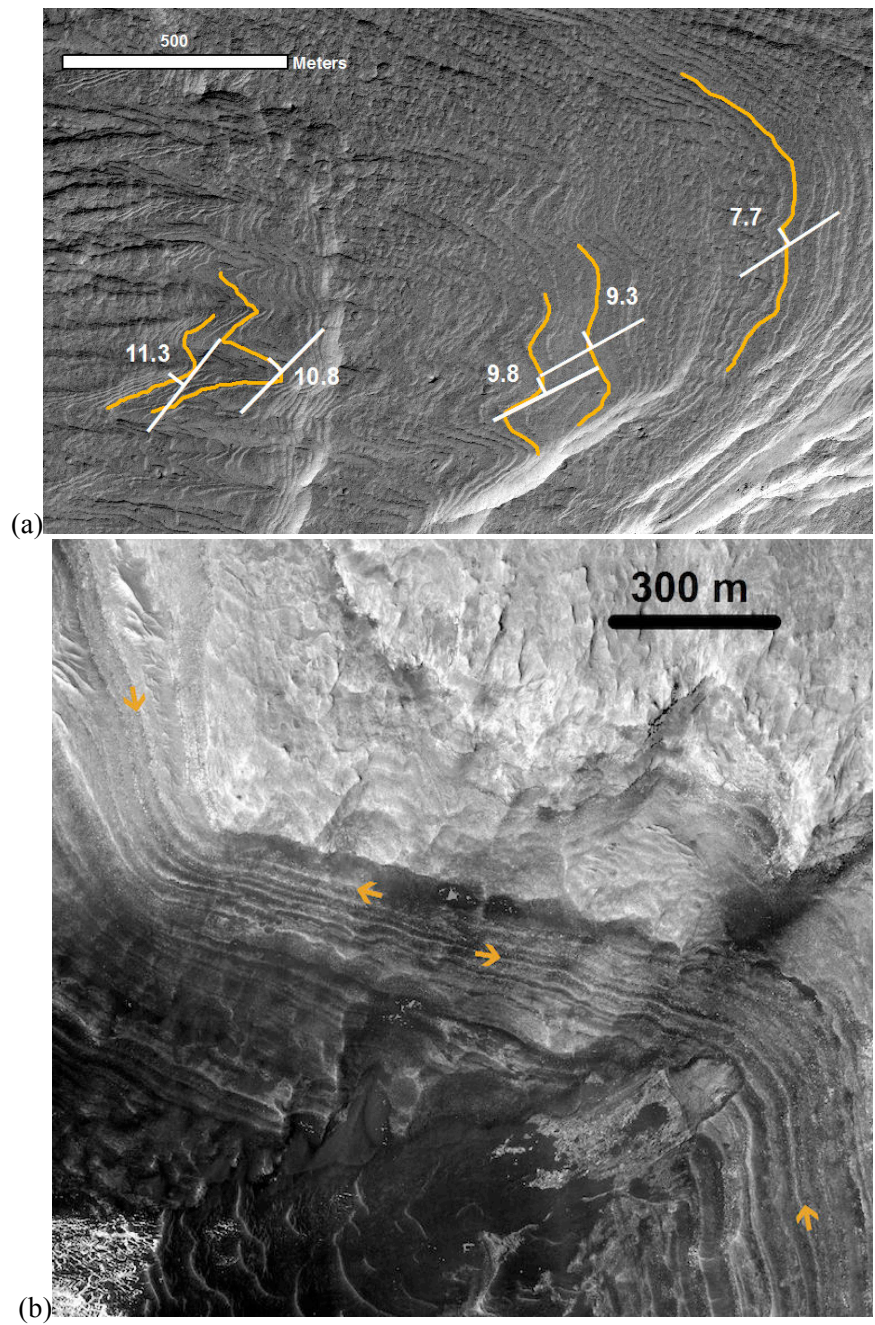

Fig. 3. (a) Traces and corresponding dips (°) for part of the ESP_017411_1711/ESP_017266_1715 stereopair. (b) Detailed trace identification for part of the reentrant canyon shown in Fig. 4a (ESP_012907_1745/ESP_013540_1745 stereopair.)

Errors in tracing a layer on a slope on an orthorectified image will produce a downslope bias in plane-fits to the trace using the corresponding DTM. Four tests show that downslope bias in our dataset does not affect our conclusions:-

(1) **Reentrant-canyon test** (Kite et al. 2013a). For the reentrant canyon at 137.2°E 5.3°S (Fig. 4a), a dominant direction of layer azimuth contrasts with a nearly complete radial rotation in dominant downslope direction. We found that layers dip in a systematic direction, typically perpendicular to local downslope. This rules out severe downslope bias.

(2) **Resolution-sensitivity test.** We compared the traces of identical layers at different image grid spacings (Fig. 5). If downslope bias affects the HiRISE layer-orientations (1m/pixel elevation model), then the same layers traced on CTX (18m/pixel DTM) will suffer a bias that is more severe. For layers in the canyon at 137.2°E 5.3°S (Fig. 5), we obtained two metrics of downslope bias (Fig. 5): (a) the angle between the best-fit plane and local topography projected onto the vertical plane parallel to steepest topographic slope, and (b) the map-plane angle between the best-fit plane and the topographic downslope. We do not find any systematic tendency for the CTX layer-orientations to be rotated downslope relative to the HiRISE layer-orientations, suggesting that the HiRISE bias is itself small.

| HiRISE Image No. | Mesa Latitude (°) | Mesa Longitude (°) | Full Dip (°) | Full Dip Direction (° CCW from E) | Cut | Dip (°) | Dip Direction (° CCW from E) | Topo. Dip (°) | Topographic Dip Direction (° CCW from E) | Full Difference from Topo. (°) | Cut Difference from Topo. (°) | **Proximity to Topo. vs. Full (°)** |
|---|---|---|---|---|---|---|---|---|---|---|---|---|
| ESP_012551_1750 | -4.852 | 137.255 | 3.03 | -150.44 | 1 | 1.09 | -105.03 | 21.67 | 21.01 | 24.67 | 22.32 | **2.34** |
| | | | | | 2 | 1.77 | -151.54 | 23.40 | -163.19 | 20.45 | 21.66 | **-1.21** |
| ESP_012551_1750 | -4.948 | 137.242 | 3.53 | 149.11 | 1 | 8.51 | 166.50 | 28.19 | 178.68 | 25.17 | 19.94 | **5.22** |
| | | | | | 2 | 2.04 | 44.01 | 29.52 | 6.61 | 32.39 | 27.93 | **4.46** |
| ESP_012551_1750 | -4.97 | 137.271 | 2.62 | 117.80 | 1 | 10.42 | 174.76 | 33.70 | 10.01 | 34.58 | 43.82 | **-9.24** |
| | | | | | 2 | 2.66 | 98.69 | 16.19 | 178.11 | 15.06 | 15.91 | **-0.85** |
| **ESP_016375_1750** | **-5.346** | **138.528** | **2.50** | **39.00** | **1** | **0.16** | **179.08** | **26.73** | **34.38** | **24.23** | **26.86** | **-2.62** |
| | | | | | **2** | **4.11** | **36.29** | **18.46** | **-153.49** | **20.91** | **22.52** | **-1.61** |
| ESP_016375_1750 | -5.335 | 138.533 | 0.62 | 89.42 | 1 | 0.73 | 127.35 | 26.73 | 34.38 | 26.38 | 26.78 | **-0.40** |
| | | | | | 2 | 4.11 | 36.29 | 18.46 | -153.49 | 18.75 | 22.52 | **-3.77** |
| ESP_012361_1685 | -11.290 | -74.68173 | 5.21 | 172.61 | 1 | 7.07 | 169.71 | 12.65 | -171.69 | 7.75 | 6.35 | **1.40** |
| | | | | | 2 | 5.21 | -169.60 | 8.10 | 25.84 | 12.78 | 13.20 | **-0.41** |
| ESP_012361_1685 | -11.220 | -74.69151 | 3.13 | -108.08 | 1 | 3.32 | -78.37 | 6.09 | -27.64 | 6.36 | 4.74 | **1.62** |
| | | | | | 2 | 3.87 | -144.70 | 10.65 | -175.29 | 9.86 | 7.57 | **2.29** |

**Table 2.** Downslope bias is shown to be small by elliptical-mesa check (Fig. 4b). "Full" refers to the entire elliptical trace. "Cut" refers to an arcuate subset of the elliptical layer, chosen to be oriented along the long-axis of the elliptical mesa (this is the worst case). The final column shows the rotation of the pole to the best-fit plane into the downslope direction, which is positive when data are consistent with downslope rotation, and negative when the data show uplslope rotation. The highlighted rows correspond to traces that are shown in Figure 4b.

(3) **Circular-mesa test.** (Fig. 4b, Table 2). In rare cases, conical topographic features show layers that can be traced in a closed loop, rather than an open curve. Because there is no obvious 'downslope direction' for closed-loop traces, the topography-induced measurement error (downslope bias) of closed-loop traces is close to zero. After tracing 7 such mesa-encircling layers, we split each elliptical trace along its ~200m-long major axis (the worst-case for downslope bias). This creates 14 test traces with a clear downslope direction and a high aspect ratio (Table 2) – again, the worst-case for downslope bias. The DTM under each trace was clipped by minimum bounding rectangle, and best-fit planes were fit to each of the traces and to the elevation data contained within the minimum bounding rectangle for that trace. The best-fit poles to the halved test cuts are consistent with zero downslope rotation. For $n = 14$, test cuts are -0.2° closer to topography on average (i.e. we find the unexpected result of upslope rotation), with a standard deviation of 3.7°, minimum of -9.2°, and maximum of 5.2°.

(4) **Geologic-control test.** HiRISE DTM layer-dip measurements made using the same technique show near-horizontal layers in areas where near-horizontal layers are expected from geological context. Specifically, near-horizontal layers have been measured from Eberswalde's delta topsets, Holden's delta topsets, and the Juventae plateau layered deposits (Irwin et al. 2015, Stack et al. 2013). These near-horizontal measurements are reported from places where the present-day erosional surface slopes steeply and so might be expected to produce large downslope bias. This geologic "control case" strongly suggests that off-horizontal Mars layer orientations are not artifacts of downslope bias, but rather geological.

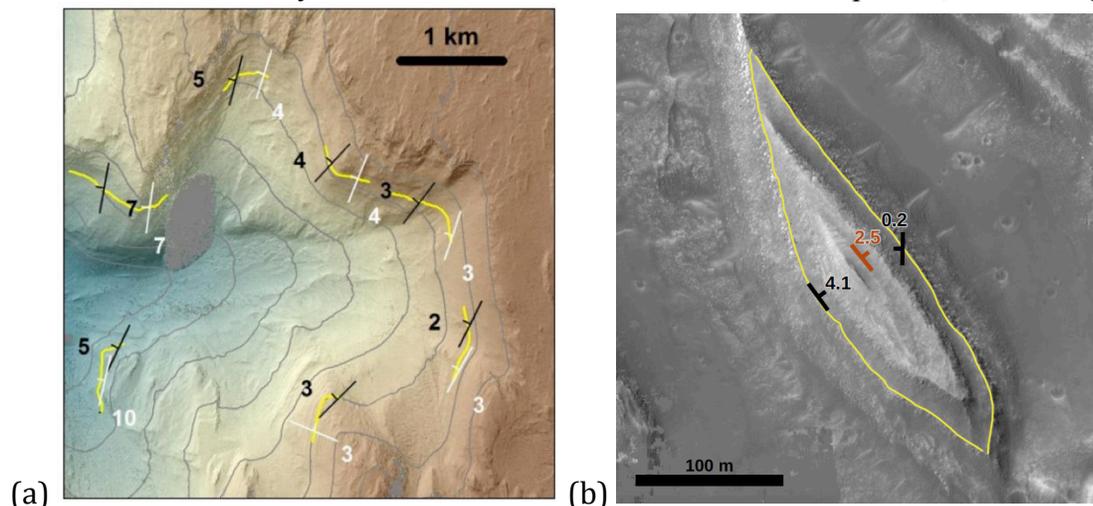

(a)          (b)

**Fig. 4.** To show that downslope bias does not affect our conclusions. (a) Examples of layers (yellow) showing similar dips whether traced on CTX DTMs (white) or HiRISE DTMs (black) in a reentrant canyon at 137.2°E 5.3°S. Gray contours show 100m topographic intervals. Brown is high. Backdrop is HiRISE DTM shaded relief. (b) Example of arcuate subsets of a layer (yellow) on a circular mesa. Mesa slope is ~23°, yet plane-fits to arcuate subsets of the layer (black symbols) show no downslope bias relative to the plane-fit to the entire elliptical trace (red symbol).

Although our checks indicate that downslope bias does not affect our conclusions, our database includes a small number (<5) of measurements where downslope bias may set the dip azimuth. These measurements all have a minimum bounding rectangle that has an aspect ratio greater than 8:1, i.e. small curvature.

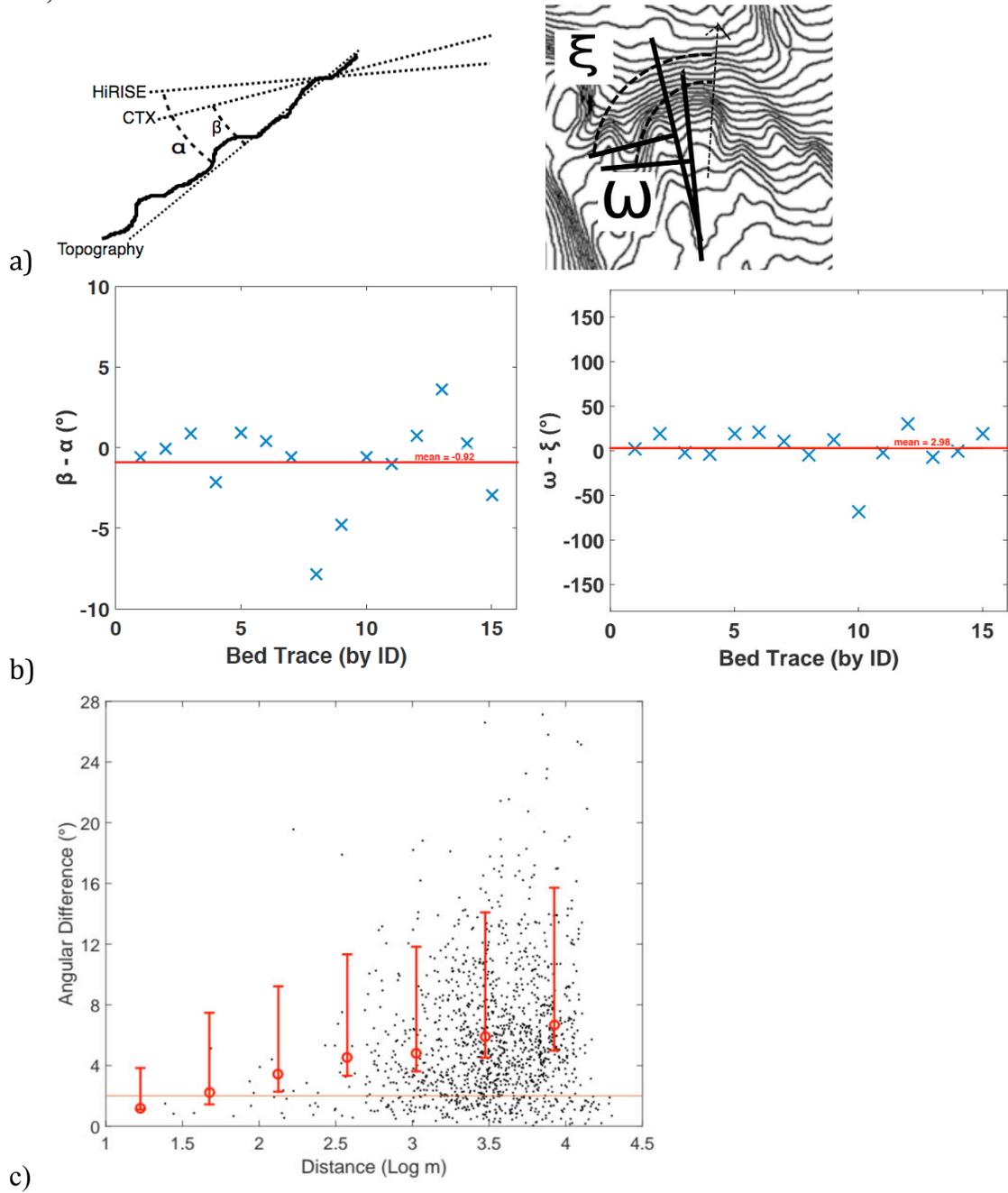

a)

b)

c)

**Fig. 5.** CTX-vs.-HiRISE layer orientation test using layer traces from the area of Fig.4a. (a) Cartoon showing how CTX-derived and HiRISE-derived layer orientations are projected onto the plane containing the downslope (topography) vector. (b) Results of HiRISE-CTX comparison. (c) Quantifying geologic noise: showing divergence between pole-fits to layers as a function of separation. Orange line shows 2° threshold. Red circles mark the mean of angular differences, binned by seperation. Red whiskers correspond to the standard deviation of the logarithms of the binned data.

We found same-worker reproducibility within error. Formal DTM precision makes a negligibly small contribution to the error. Consistency between measurements by workers in the University of Chicago and Johns Hopkins University labs using the same procedure was demonstrated. Between-lab reproducibility for poles-to-layer-planes in NW Gale (JHU vs. Chicago) was 1.3° on average (standard deviation 1.0°, worst-case 3.7°, $n = 17$), which is less than our error bars. Outward dips at Mt. Sharp / Aeolis Mons have been independently confirmed by Fraeman et al. (2013), Le Deit et al. (2013), and Stack et al. (2013).

Within-measurement errors (the residuals of measured points around the best-fit plane) were quantified using the method of Lewis (2009), which is a conservative approximation to a 95% regression-error estimate. Points with >2° error were rejected. The mean pole error in the whole database is 0.98°. Same-worker reproducibility averaged 1.3°, with a standard deviation of 1.0°. The same-worker reproducibility check layers were chosen to systematically span a range from smallest to largest $\Delta Z$, where $\Delta Z$ is the absolute range of elevation values. We did not find any tendency for reproducibility to get worse with decreasing $\Delta Z$. However, small-$\Delta Z$ traces remain sensitive to small-scale geologic variation (e.g. fractures, boulders), so caution is warranted in interpretation of individual traces with $\Delta Z < 3.5$ m.

To quantify between-measurement variations (geologic noise), we plotted (for each DTM) pairwise angular differences between the poles-to-layer-planes as a function of the pairwise separation between median {x,y} positions of individual traces (Fig. 5c). We found that the pairwise differences are well-fit by a line that increases log-linearly with separation, and intersects 180 m at ~4° (Fig. 5c). Within-DTM differences in layer orientation can greatly exceed our error bars, and so are likely real (geological). These layer-orientation differences could be primary depositional features, or the result of short-wavelength postdepositional tilting.

Together, these tests show that our measurements are accurate and reproducible and that downslope bias does not affect our conclusions. We cannot exclude a selection bias (layers that dip close to slope will have corrugated outcrops that are easier to measure). However, our measurements cover many mounds and a broad range of stratigraphic elevation, minimizing this effect. For the purpose of understanding mound build-up, within-DTM scatter in the measurements (km-wavelength geologic noise) sets the practical limit on interpretation - not measurement precision or accuracy.

Results are given in Figs. 6-9 and section 3.

### 2.3. Fitting of stratigraphic surfaces interpreted as erosional unconformities.
We traced stratigraphic surfaces (interpreted by previous workers as erosional unconformities) in W. Candor, Ophir, and Gale (Anderson & Bell 2010, Thomson et al. 2011, Le Deit et al. 2013, Lucchitta 2015). We interpret the traces as unconformities on the basis of a sharp break in tone, erosional or layering style, crater density, or slope, at a stratigraphic level that, in at least one location, corresponds to an unconformity (shown by buried craters, or by truncated layers) (e.g. Fig. 9). In none of these cases is definitive unconformity mapping possible using CTX data alone, and complete HiRISE coverage is not available. In W. Candor and Gale, we believe that the traces do correspond to major unconformities (e.g. Fig. 9), and that our traces follow a stratigraphic surface sufficiently closely to determine the qualitative paleotopography (dome, trough, saddle, or roughly flat), and to put lower bounds on isochore measurements. Next, we made use of DTMs constructed using CTX stereo data (for Gale) or using MOLA data (for W. Candor). For segments of the trace where we were confident about the location of the unconformity, we calculated the total relief (max. elevation – min. elevation) of the unconformity trace. Next, the digitized points were interpolated to form unconformity surfaces using (i) inverse distance weighting, (ii) planar interpolation, and (iii) quadratic global polynomial interpolation. In principle, the interpolation procedure is subject to a dome bias that is analogous to the downslope bias in layer-orientation fits. In practice however, the >km total relief of the unconformity traces means that any such bias is unimportant for the purpose of determining the best-fit shape of the stratigraphic surface. Following interpolation, we subtracted these surfaces from present-day topography to create thickness contours (isochores) for the material above the within-mound unconformities. Results are given in Figs. 10-12 and section 4.1.

### 2.4. Identification of draped landslides.
We identified mass-wasting units (flows, slides, spreads, falls and topples), which we refer to as "landslides", using THEMIS and CTX images. Comparison with a preliminary U.S. Geological Survey geologic map of the Central Valles Marineris (Fortezzo et al. 2016) shows that our identifications of mass-wasting zones agree. We additionally looked for locations where undeformed layered materials superposed the source zones of the landslides, indicating layered material deposition after moat formation (Anderson & Bell 2010, Okubo 2014, Neuffer & Schultz 2006). Results are given in Figs. 13-14 and section 4.2.

# 3. Layer-orientation results.

### 3.1. Overview.
Among our measurements (308 layer dips extracted from 30 DTMs), most strata within VM and Gale's mound were found to dip away from mound crests (Fig. 6b). For layers above the mound base, the dip azimuth of 87% of the measured layers falls within 90° of the vector directly away from the nearest mound crest (mound centroid for Gale); 57% are aligned within 45°. This tendency is equally strong in Gale's mound ($n = 126$) and VM ($n = 182$) (Fig. 7d). The median dip of the measurements in our database (5°) corresponds (for an 80-km wide mound) to 3.5 km of relief on a stratigraphic surface. Indeed, canyons carved into Gale's mound show easily-observable relief of 500m on individual layers.

Lowermost strata (≤0.5 km above the interpolated basal surface), which will soon be visited by the MSL rover, still dip preferentially away from mound crests (Fig. 6c). This structural consistency with elevation contrasts with the mineralogical variability observed at Gale and elsewhere (Milliken 2010). Sulfate detections specifically correspond to outward-dipping layers at Ganges Mensa, Melas Mensa, and Gale's mound (Chojnacki & Hynek 2008, Fueten et al. 2014; Fig. 8), as well as for Hebes and Candor Mensae (Schmidt 2016, Jackson et al. 2011, Fueten et al. 2014). Furthermore, draping layers in SW Melas Chasma show sulfate signatures (Weitz et al. 2015), suggesting that sulfate-bearing rocks on Mars can form at primary depositional angles that are far from horizontal. Therefore, preferentially-outward dips are not restricted to the spectrally-bland, capping 'rhythmite' facies identified in many sedimentary deposits on Mars, including the uppermost Gale strata (Grotzinger & Milliken 2012, Lewis & Aharonson 2014). Although the rhythmite facies lies topographically above the northern rim of Gale crater, its induration still suggests cementation involving liquid water (Lewis et al. 2008).

Just as the mineralogical transitions up-section do not correspond to the end of surface liquid water on Mars, our measurements further suggest that they need not be accompanied by a change in the physical process of deposition.

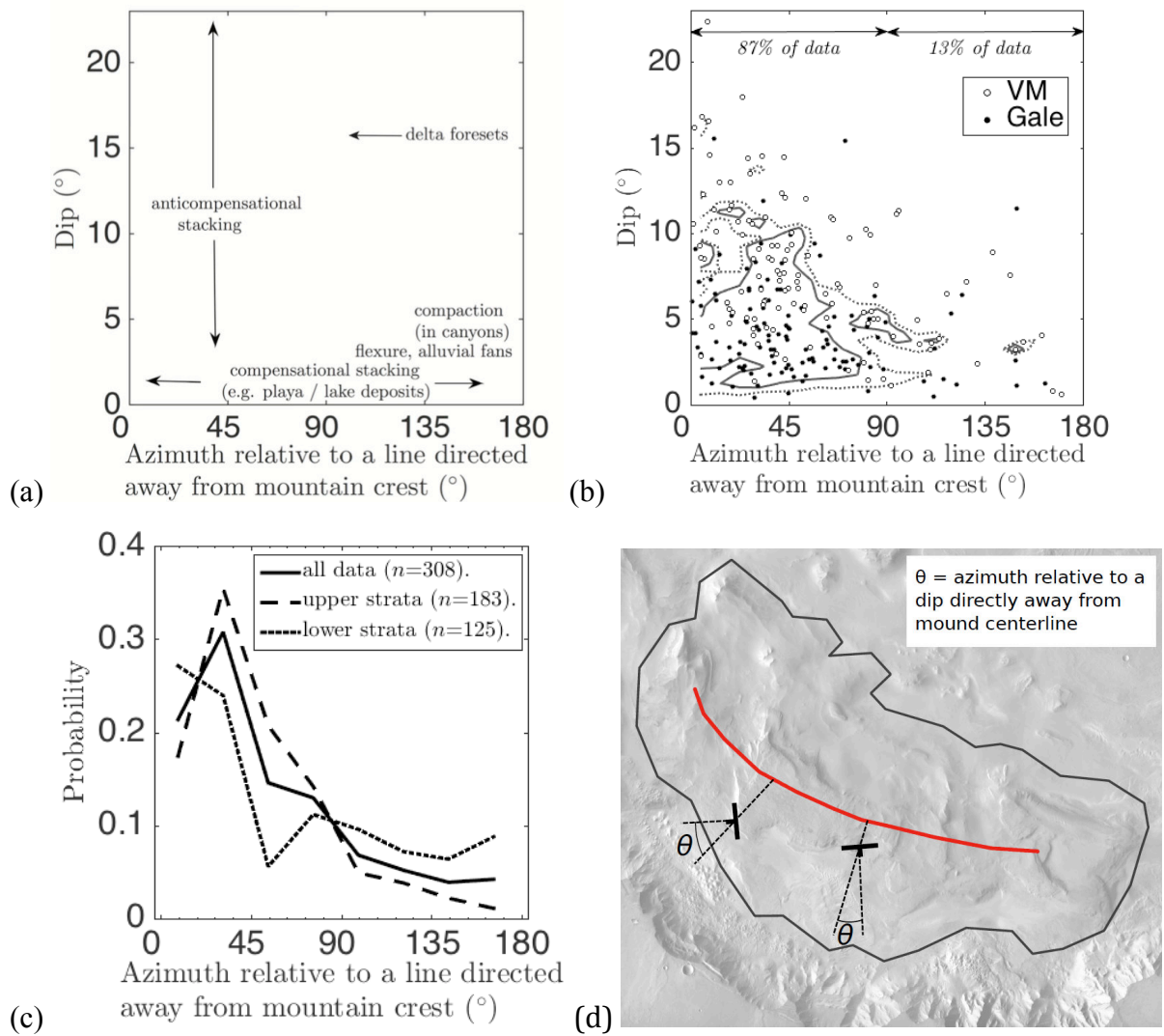

(a)

(b)

(c)

(d)

**Fig. 6.** Layer-orientations summary: (a) Expectations for primary depositional orientation (Leeder et al. 2011, Moore & Howard 2005, Davis 2007, Kite et al. 2013a, Grotzinger et al. 2015). "Delta foresets" refers to a container-wall sediment source. (b) Results, for measurements above the interpolated basal surfaces of the mounds. Solid and dashed contours enclose 50% and 68% of data, respectively, after accounting for heteroskedastic error. Marginalizing over dip, 87% of the azimuth data lie within 90° of a line directed away from mountain crest. (c) Distribution of layer orientations relative to elevation above interpolated basal surface of mound (lower strata are ≤0.5 km above mound base; upper strata are >0.5 km above mound base; 22.5° bins).

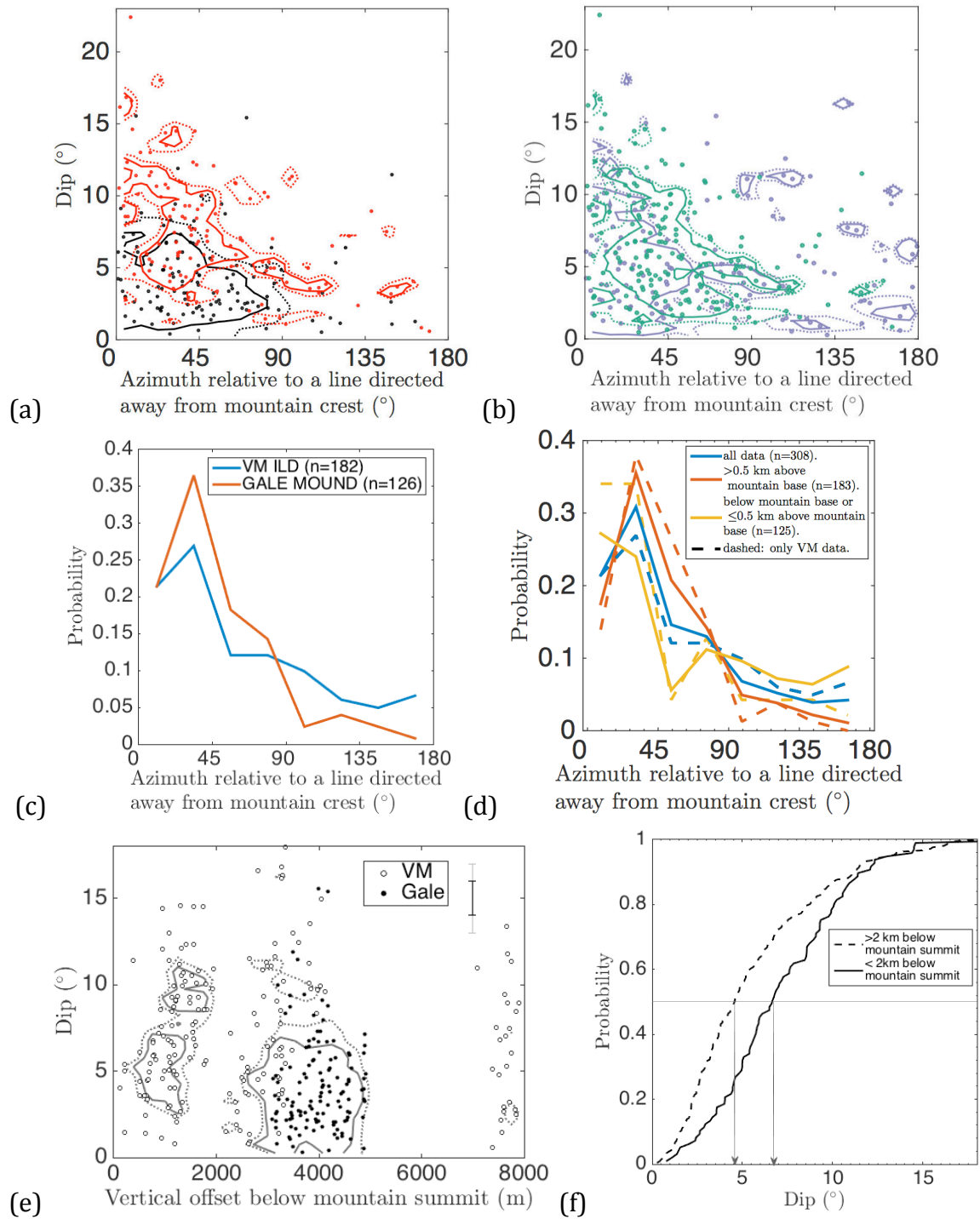

**Fig. 7.** Layer-orientation details: (a) Data from Gale (black) compared to data from VM (red). Solid and dashed contours enclose 50% and 68% of measurements, respectively. (b) Layer-orientations ≤0.5 km elevation above interpolated basal surface (purple) compared to layer-orientations >0.5 km above interpolated basal surface (green). Solid and dashed contours enclose 50% and 68% of measurements, respectively (Compare Fig. 6c). (c) Dip azimuth of all layers in Gale compared to dip azimuth of all layers in VM. (d) Distribution of layer orientations relative to elevation above interpolated basal surface of mound (22.5° bins). (e) Distribution of layer dip

with vertical offset below mound summit (filled symbols for Gale data, open symbols for VM data). Gray line shows maximum error of included data points, and black line shows mean error of included data points. Solid and dashed contours enclose 50% and 68% of measurements, respectively. (f) Cumulative probability distribution function for dip amplitude, categorized by distance below mound summit.

Median dip >2 km below mound summit is 4.7° deg ($n$ = 216), less than median dip ≤2 km below mound summit (7.0°, $n$ = 92) (Fig. 7). Similarly, layers >1.5 km above the interpolated basal surface ($n$ = 99) dip more steeply (median 7.5°) than layers ≤1.5 km above the interpolated basal surface ($n$ = 209, median dip 4.5°). Gale data are shallow-dipping and >3km below mound summit, and removal of Gale data (or removal of VM data) would remove the dip-versus-elevation correlation in our database.

The tendency for layers above mound base surface to dip away from the center of the mounds is insensitive to the error threshold (cutoff) beyond which data are discarded. Our nominal cutoff of 2° gives 87% of layers dipping away from the mound center. A cutoff of 1° gives 84% of layers dipping away from the mound center. Accepting all measurements, with no cutoff, yields ~10% more layer-traces but no change in the percentage of layers that dip away from the mound center (87%).

The data indicate a strong preference for layers to be oriented away from mound centerlines.

### 3.2. Seven of the eight mounds investigated individually exhibit the outward dips predicted by Kite et al. (2013a).

Mound-by-mound analysis shows that seven of these eight mounds studied in this paper individually exhibit the outward dips predicted by Kite et al. (2013a) – Gale's mound, plus Ceti, Ophir, Melas, Ganges, Nia, and Juventae Mensae (Fig. 8). For each mound, we visually inspected the intersections of layers in our CTX orthophotos with contour lines generated using our mound-spanning CTX DTM mosaics and confirmed that these structure contours are qualitatively consistent with the patterns described below using HiRISE data.

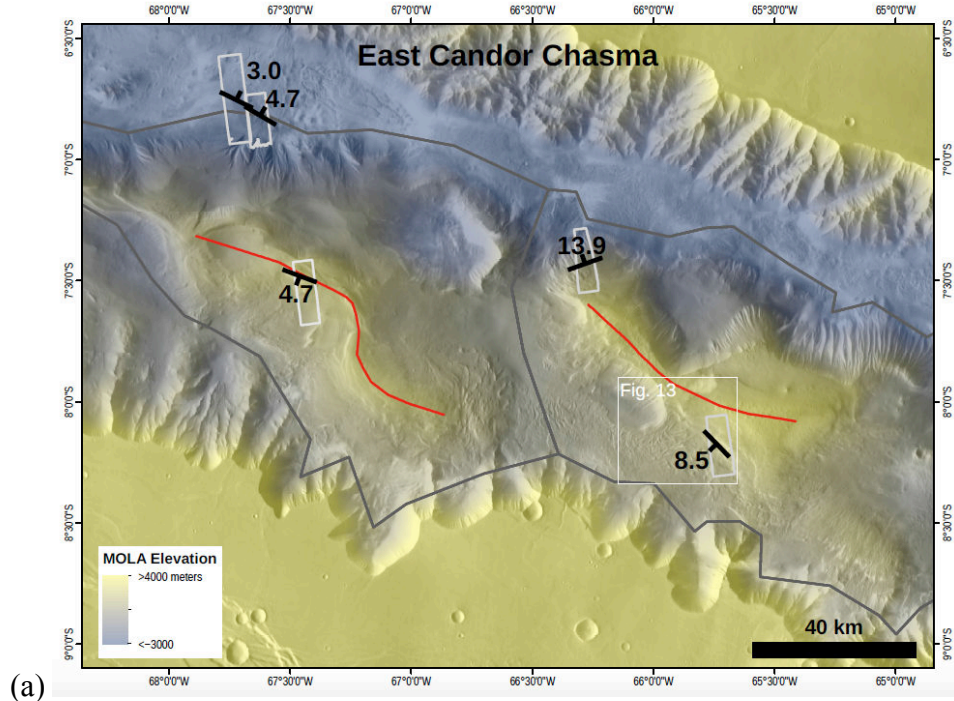

(a)

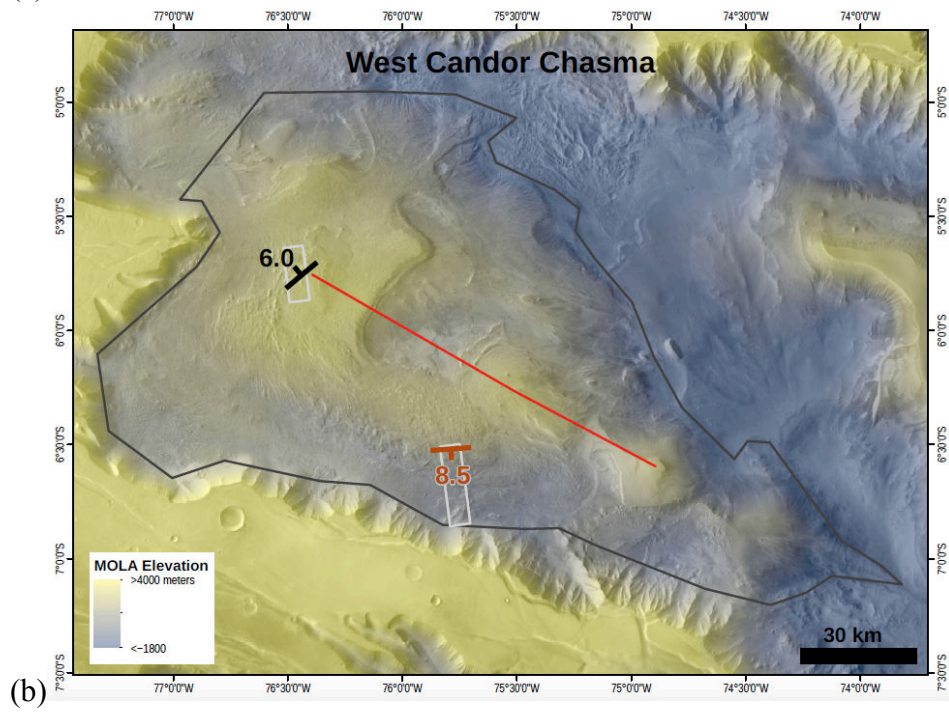

(b)

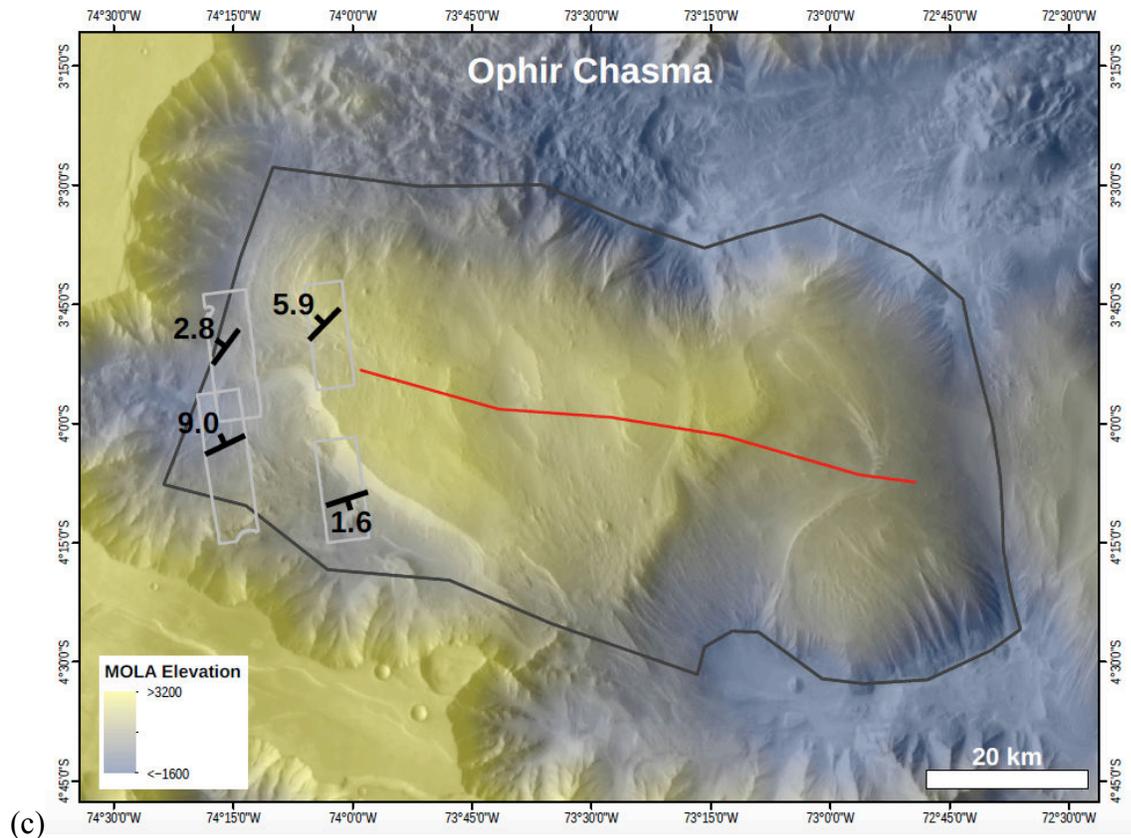

(c)

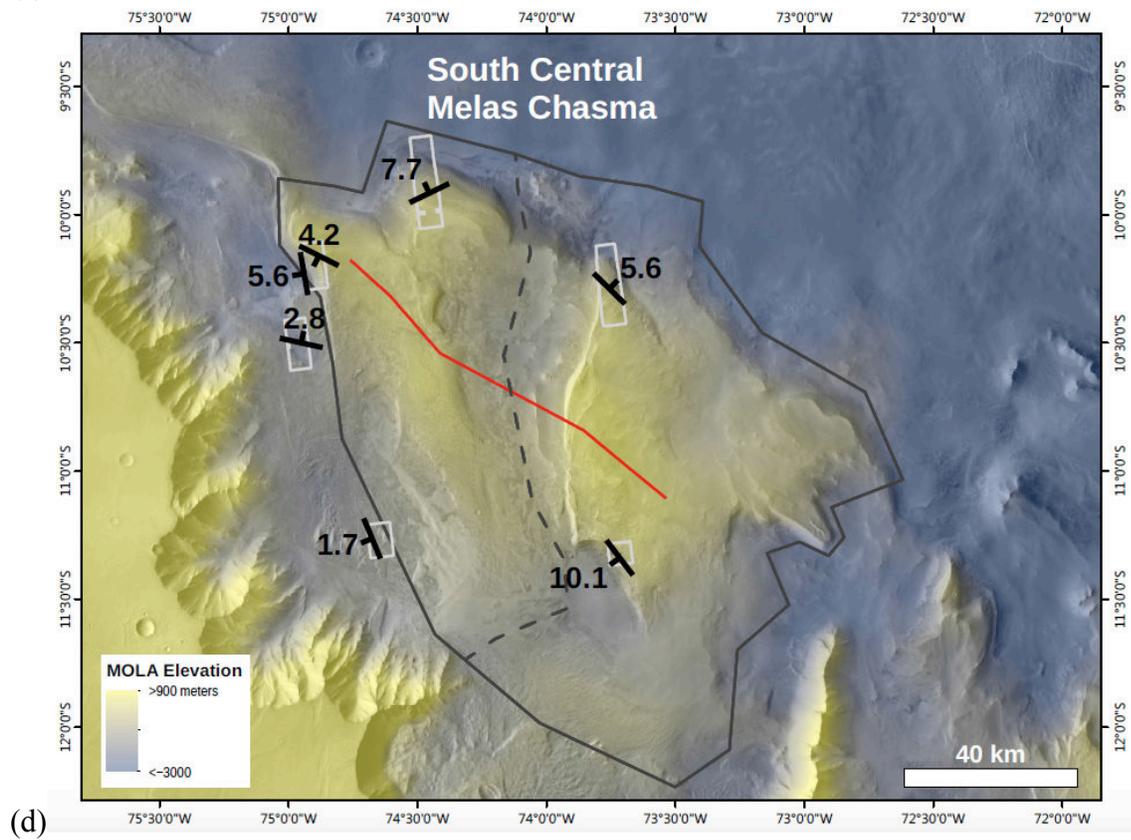

(d)

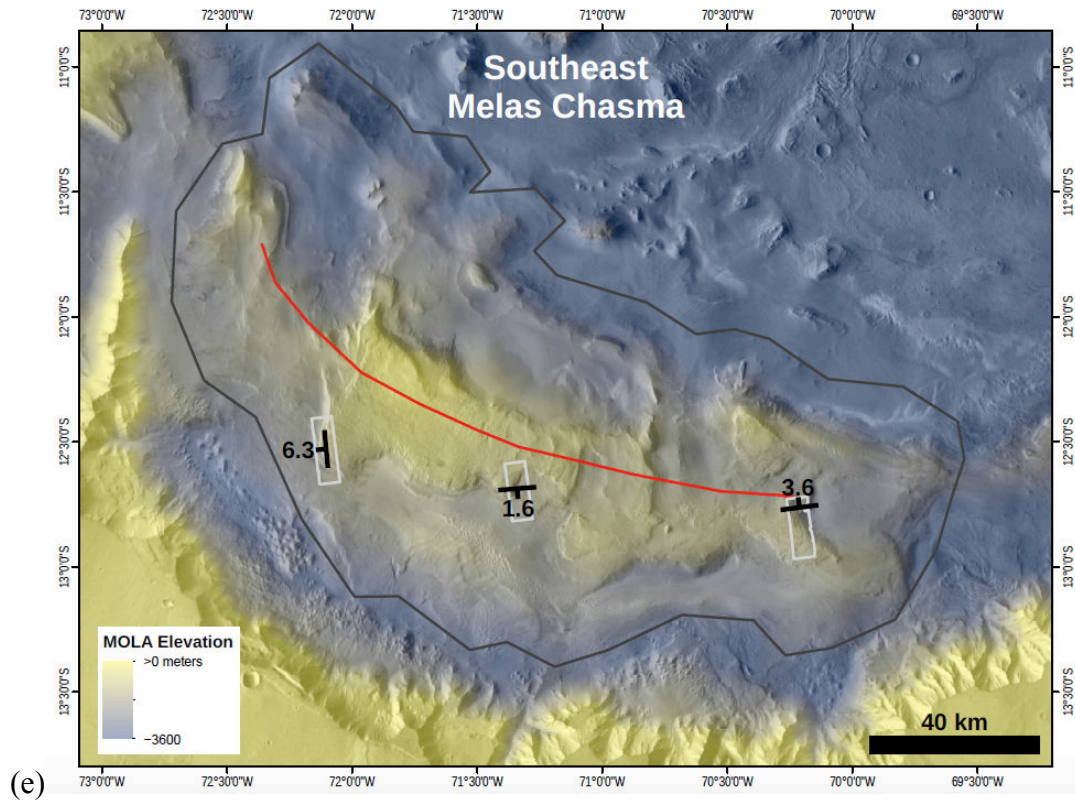

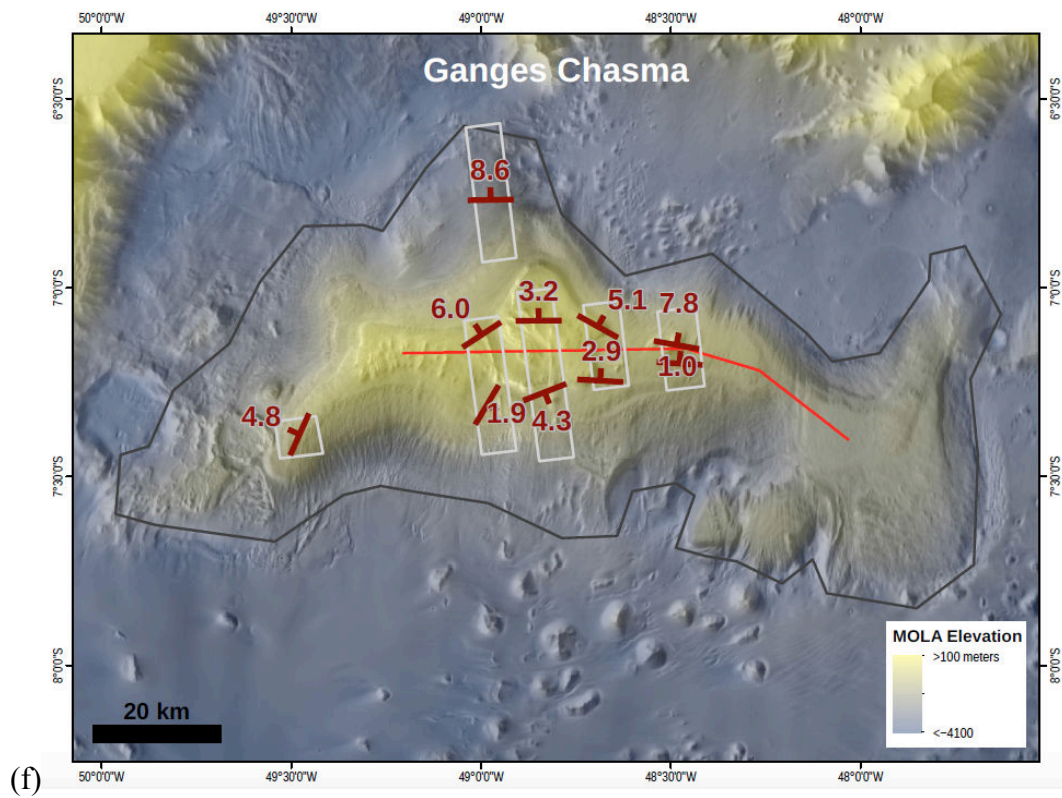

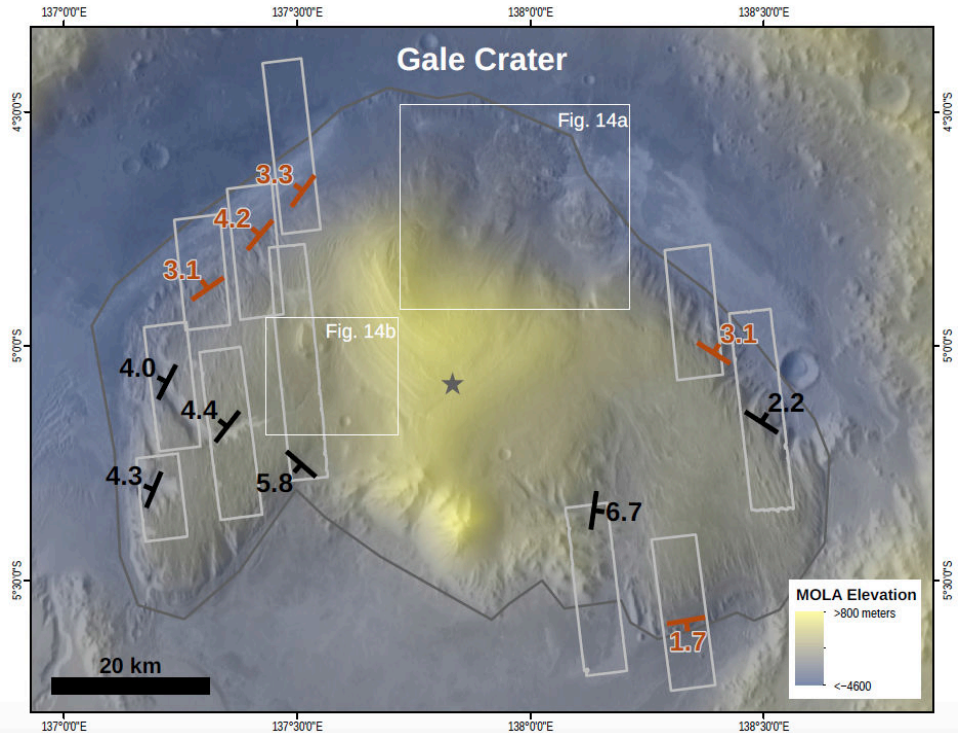

(g)

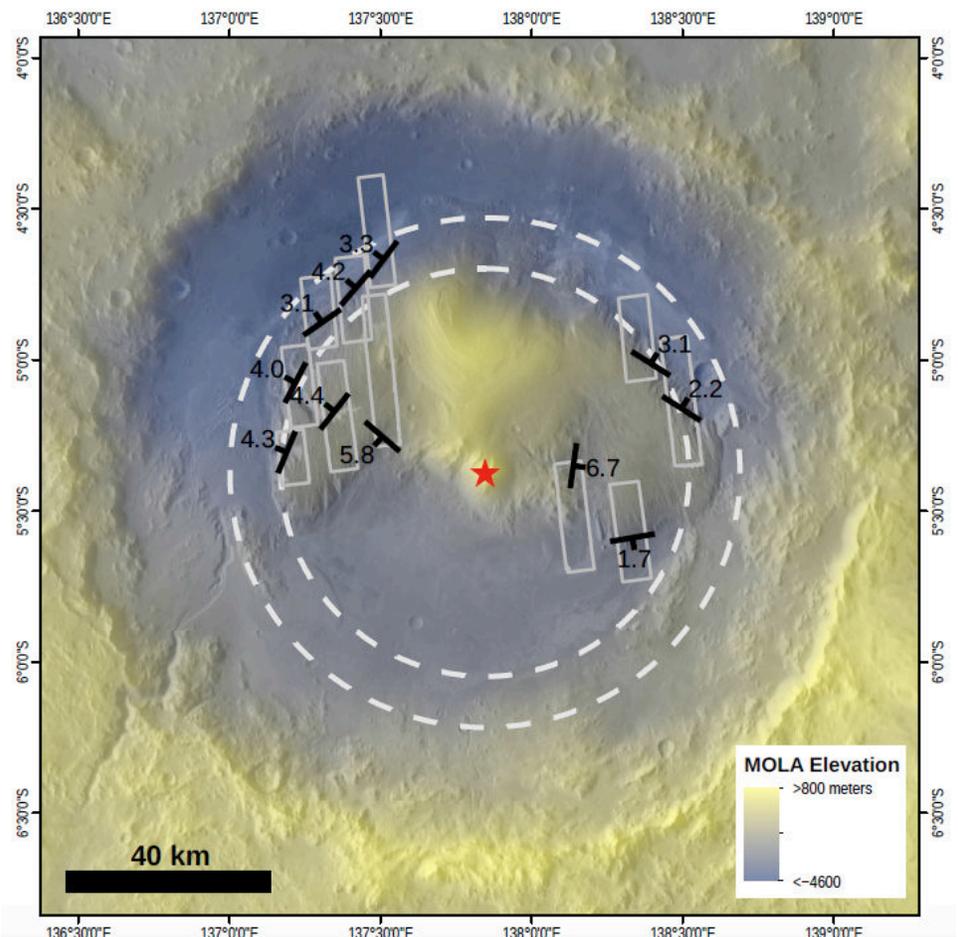

(h)

**Fig. 8.** Mound-by-mound layer-dip data (dips in degrees). Color scale is clipped at high and low elevations in order to emphasize mound topography. Light gray shows the outlines of the HiRISE orthoimages/DTMs. Dark gray lines enclose the topographically-defined mounds. Red lines show the mound crest-lines. Strike-dip symbols and labels indicate average orientations of all layers traced on the corresponding orthoimage/DTM. **(a)** East Candor Chasma contains Nia Mensa (west) and Juventae Mensa (east). White rectangle shows the location of the draped landslide shown in Fig. 13. **(b).** West Candor Chasma contains Ceti Mensa (drawn to include Nia Tholus). Crest-line is drawn across a late-stage erosional window in central Ceti Mensa. See also Fueten et al. (2006) and Murchie et al. (2009a, 2009b). **(c).** Ophir Chasma contains Ophir Mensa. Crest-line is drawn to cut across a late-stage erosional window in the east of the mound. See also Wendt et al. (2011). **(d).** South-Central Melas Chasma contains Melas Mensa. Crest-line is drawn to crosscut a topographic low that is interpreted as an erosional trough. An alternative scenario, in which Melas Mensa is in fact two mensae, is indicated by the dashed line. **(e).** South East Melas Chasma contains Coprates Mensa. Crestline is drawn to cut across some small troughs. **(f).** Ganges Mensa (data from Hore 2015). For DTMs that straddle the mound centerline, we plot the average for data north of the centerline separately from the average for data south of the centerline. **(g).** Gale crater contains Mt. Sharp / Aeolis Mons. Gray star shows the centroid of Gale's mound. Main canyon of Gale's mound is to the W. White rectangles show location of draped landslide and draped canyon in Fig. 14a and 14b, respectively. **(h).** Showing Gale data in relation to Gale crater. Range rings (white dashed lines) show distance of candidate peak-ring (Allen et al. 2014) from Gale's central peak (red star).

*East Candor Chasma* (Fig. 8a). E Candor contains the tallest sedimentary rock mounds on Mars, Juventae Mensa and Nia Mensa. Five HiRISE DTMs were obtained, whose mean dips ($n = 46$) systematically point away from the present-day mound crests (Fig. 8a). One DTM (ESP_034896_1725/ESP_036542_1725) has only 2 traces within the 2° error threshold. Near the base of Nia Mensa, an arcuate feature has been interpreted as a delta (Le Deit et al. 2008): in our layer-trace database, this feature shows northward dips that lack the fanning-out dip-directions expected of a delta. Nia Mensa and Juventae Mensa are dusty, and we are not aware of published sulfate detections there (Roach 2009).

*West Candor Chasma* (Fig. 8b). Measured dips within Ceti Mensa are outwards. Our Ceti Mensa observations support Okubo's (Okubo et al. 2008, Okubo 2010, Okubo 2014) interpretation of a paleo-moat. The red point in Fig. 8b was calculated by taking the average of the 210 dips reported by Okubo (2014) from the northernmost (highest in elevation) outcrop of the $CeM_k$ unit as defined in Okubo (2014). Sulfate minerals are found at levels stratigraphically equivalent to many of the outward dips (Gendrin et al. 2005, Mangold et al. 2008, Murchie et al. 2009). Our one additional DTM W of the mound centerline has 3 good traces, showing generally W-directed dips.

*Ophir Chasma* (Fig. 8c). Four HiRISE DTMs ($n = 48$), all from the western end of the Ophir Mensa mound, show dips that are directed away from the mound crest except for the lowest-elevation DTM, which shows dip directions that parallel the mound crest. The DTM marked "2.8°" shows layers that drape the lowest part of the canyon wall in the W of the DTM, and layers that dip steeply towards the canyon wall in the E of the DTM. Dust on Ophir Mensa complicates spectroscopy. We know of one kieserite detection on Ophir Mensa at the level of our measurements (Gendrin et al. 2005, Chojnacki & Hynek 2008).

*South Central Melas Chasma* (Fig. 8d). SC Melas Chasma hosts Melas Mensa, an ~3km-high mound with 6 HiRISE DTMs (*n* = 60). Layers dip away from the mound centerline around the mound. Sulfate minerals are found at levels stratigraphically equivalent to many of the outward dips (Gendrin et al. 2005, Chojnacki & Hynek 2008). However, DTMs of sedimentary rock layers at the base of the mound show more variable layer-orientations, including some layers dipping back toward the mound. Melas Mensa has a N-S aligned medial trough. If this trough is used to divide the mound into two mounds, then traces from two elevation maps are most strongly affected (ESP_012361_1685_ESP_012572_1685 and PSP_005953_1695_PSP_002630_1695), totaling 21 traced layers. Under likely measures of two central ridge features for each mound considered separately, 9 of these traces would be oriented more directly away from central ridges, and 13 would be oriented less directly. Therefore, our conclusion is not affected by whether Melas Mensa is considered as 1 or 2 mountains.

*South-East Melas Chasma* (Fig 8e). SE Melas Chasma hosts Coprates Mensa, a relatively low ~2km mound with layer orientations that do not match those predicted by Kite et al. (2013a). Two of our three DTMs (*n* = 23) show layer orientations that are variable, but average out to dips that are parallel to the mound-crest; the remaining DTM shows dips that slope back toward the mound crest. Sulfates exist at the level of some of our measurements in the E of Coprates Mensa (Gendrin et al. 2005, Chojnacki & Hynek, 2008).

*Ganges Chasma* (Fig. 8f). A comprehensive (6 DTMs) study of Ganges Mensa (Hore 2015) shows systematic outwards dips (Fig. 8f). Hore (2015) does not provide error bars, so we do not include Ganges data in Figs. 6-7. Sulfates are common in Ganges Mensa (Chojnacki & Hynek 2008), including at the stratigraphic level of our measurements.

*Gale crater* (Fig. 8g, 8h). Our database for Gale's mound, Mt. Sharp / Aeolis Mons (*n* = 126) includes new measurements for 6 DTMs. We combine these with layer orientations from 5 DTMs presented in Kite et al. (2013a). Layers dip systematically away from the mound center, including up the main canyon of Mt. Sharp / Aeolis Mons and close to the center of the mound. However, we did not find many traceable layers in HiRISE stereopairs close to the center of the mound, and the 2 DTMs closest to the center of the mound each have only 3 traces within the 2° error threshold. The observed persistence of outward layer dips up the main canyon of Gale's mound rules out the hypothesis that a peak ring is solely responsible for the layer-orientations. Arcuate mounds between 40 km and 50 km from Gale's central peak may be eroded remnants of a peak ring (Allen et al. 2014). If this central ring persists underneath Gale's mound, then it might affect layer orientations locally. We did not find clear evidence for a peak ring effect on layer orientations; it is possible that further analysis of HiRISE DTMs might turn up such evidence. Whether or not a "peak ring effect" is detectable in the orientation of some of the layers of Mt. Sharp / Aeolis Mons, the outward layer dips we observe occur at a wide range of distances from Gale's central peak – too wide a range for a peak ring to explain the outward dips. Because Gale's central peak is volumetrically negligible compared to the volume of Mt. Sharp's lower unit, and is visibly intact, erosion of Gale' central peak cannot account for the deposits contained within Aeolis Mons / Mt. Sharp's lower unit.

Hebes Mensa (not shown) shows systematic outward dips, and sulfate detections, but a flat unconformity (Jackson et al. 2011, Schmidt et al. 2015, Schmidt 2016).

# 4. Stratigraphic unconformities and draped landslides.

a)                                                  b)

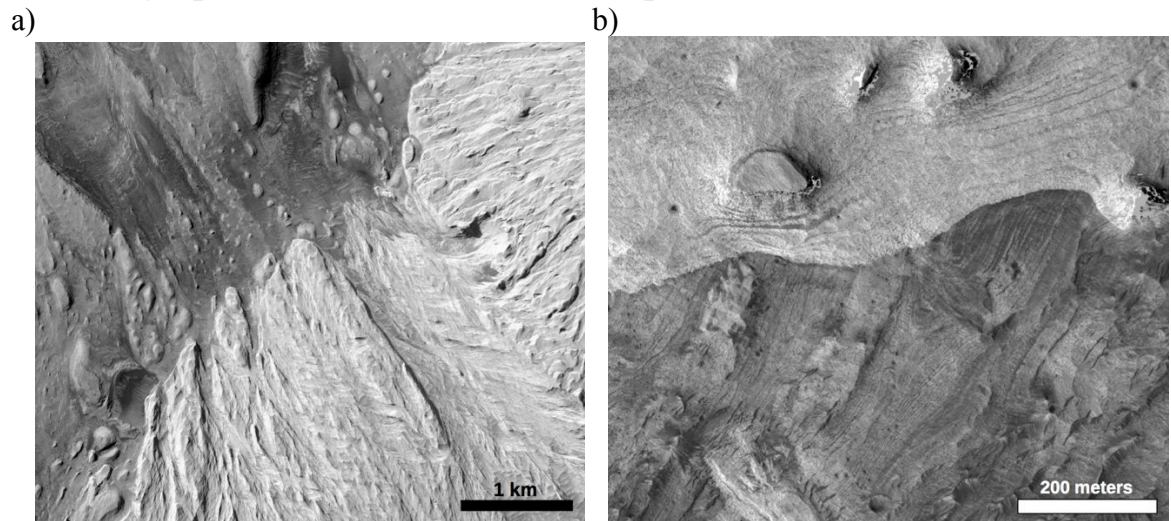

**Fig. 9.** Examples of unconformities. (a) HiRISE snapshot of a mound-spanning unconformity in a within-crater mound (Gale's mound, near 4.83S 137.41E). Note embedded crater in bottom left. **(b)** HiRISE snapshot of a large unconformity in a within-canyon mound (Ceti Mensa, near 5.80S 76.47W).

## 4.1. Stratigraphic unconformities.

We analyzed the major unconformities reported at Gale and W Candor (Anderson & Bell 2010, Lucchitta 2015, Thomson et al. 2011, Le Deit et al. 2013) (Figs. 9-12). In every case (Table 3), present-day exposures of these surfaces show large (1-4 km) relief (Malin & Edgett 2000, Fueten et al. 2014). Using procedures described in Section 2.3, we identify and interpolate these stratigraphic horizons across each deposit. The interpolated unconformity surfaces dip steeply (Thomson et al. 2011), analogous to the modern mound forms, and consistent with past wind erosion (Heermance et al. 2013). Interpolated surfaces typically define paleo-domes within the interior of each mound (Table 3). Paleo-dome summits are usually close to modern topographic highs (Fig. 9). Furthermore, isochores show preferential deposition near paleo-dome summits (Figs. 9-12). These paleo-domes, defined by unconformable surfaces deep within the stratigraphy, strongly suggest the existence of moats during the interval of deposition (i.e., anti-compensational stacking).

Our data suggest that that the mound-spanning unconformities truncate underlying layers, usually dip toward the canyon edge or crater rim, and are draped by parallel layers (Anderson & Bell 2010, Holt et al. 2010, Okubo 2014). Draping implies that post-unconformity sediments were wind-transported (similar to Holt et al. 2010). Water-transported sediments would onlap the paleo-dome. We looked for, but did not find, evidence for onlap. We are not aware of basin-scale unconformities of this type on Earth.

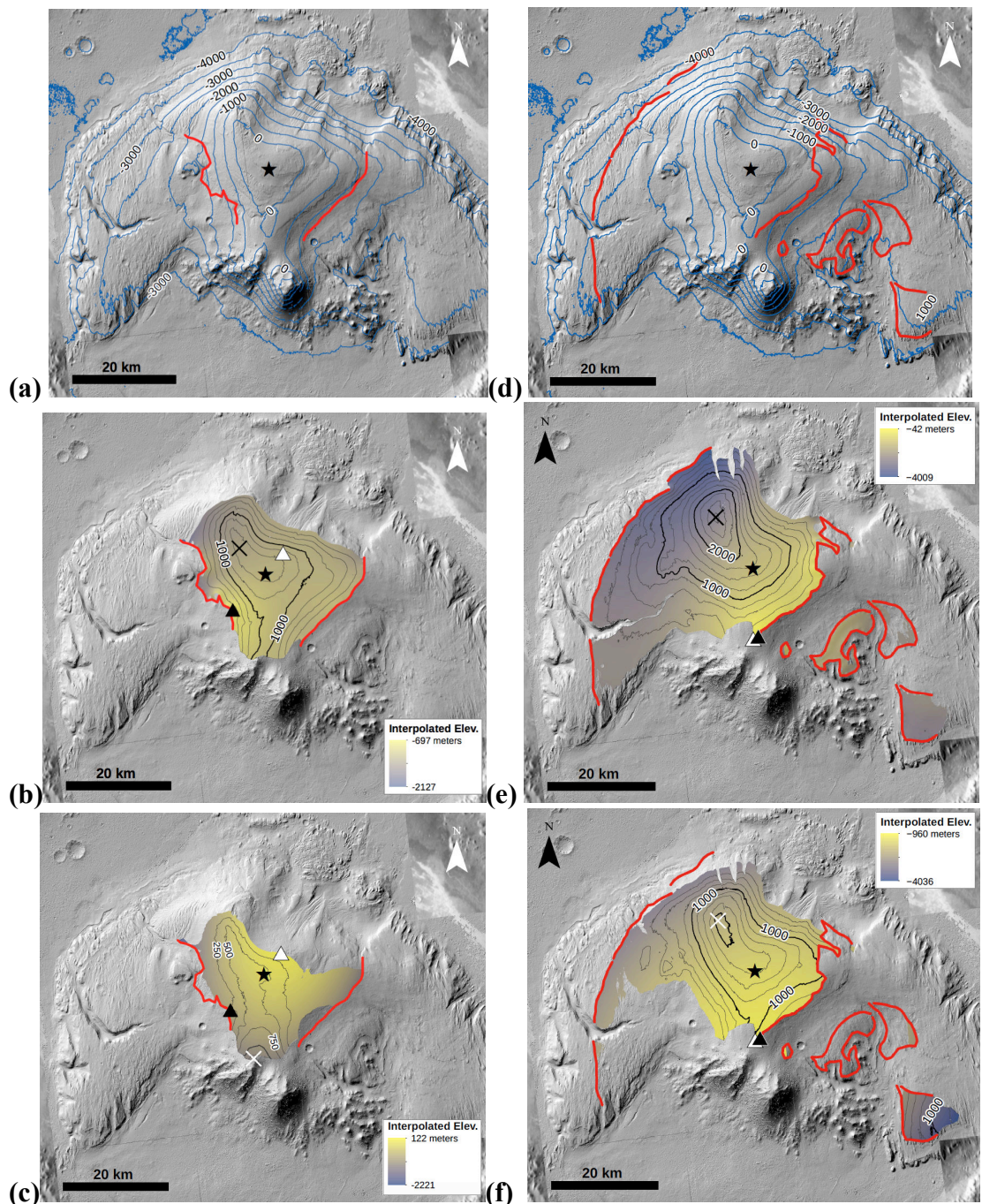

**Fig. 10.** Paleo-domes within Mars mounds. (a) Topography (blue contours) for Gale's mound (Gale is 155 km diameter). Red lines show the trace of stratigraphic surface interpreted as unconformity by Anderson & Bell (2010), which has >1 km of relief. ★=sedimentary-mound summit. (b, c) Colors show paleo-topography of Gale's mound, interpolated using (b) inverse-distance weighting and (c) quadratic polynomial interpolation. Black contours show isochores for late-deposited material. ▲,△=high points of unconformity surfaces (filled for IDW-interpolated, open for quadratic-interpolated). ×=locations of maximum thickness for upper units. (d, e, f): As (a-c), but for stratigraphic surface interpreted as unconformity from Thomson et al. (2011). Background is shaded relief of CTX DTM mosaic.

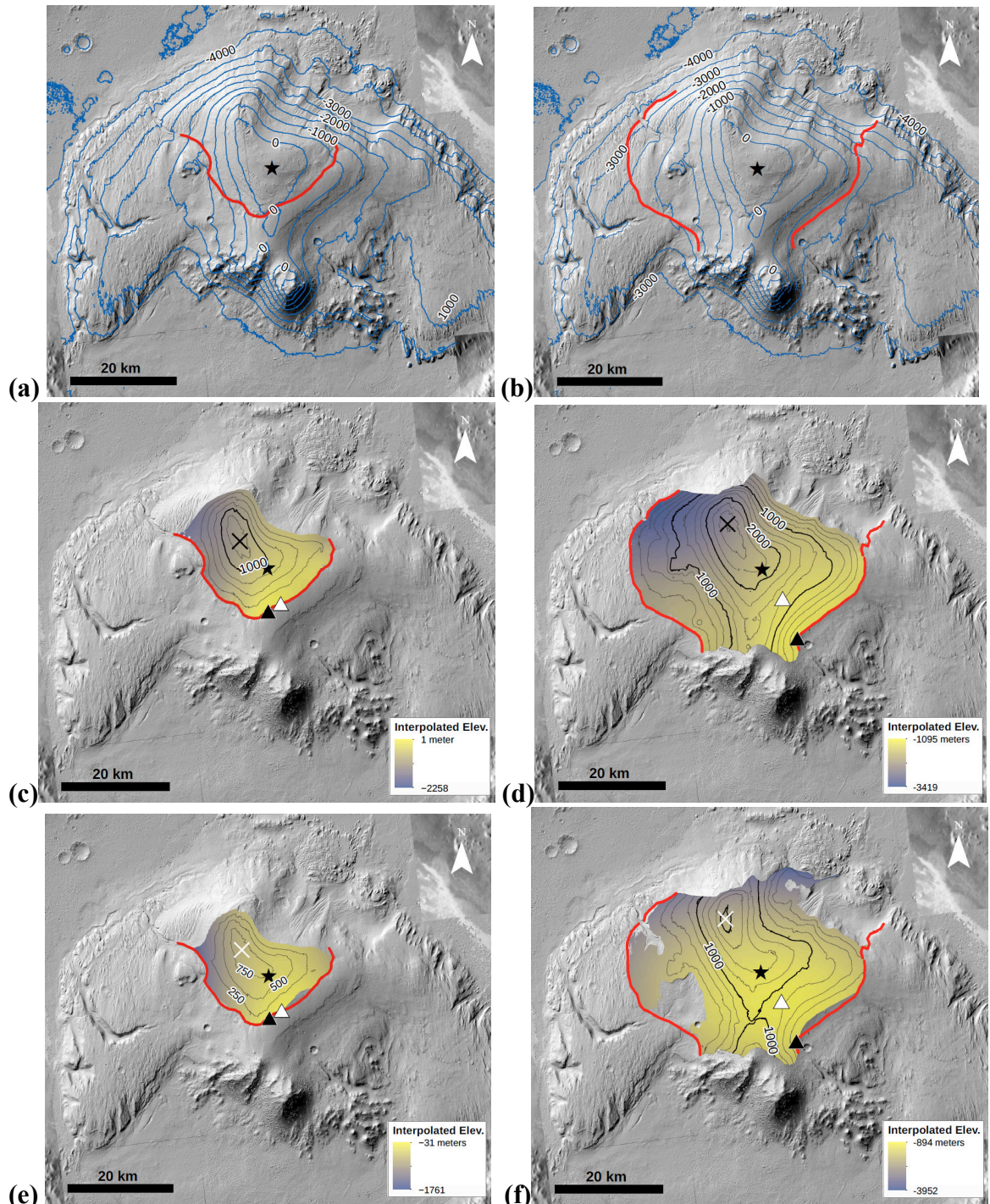

**Fig. 11.** Paleo-domes within Mars mounds. (a) Topography (blue contours) for Gale's mound (Gale is 155 km diameter). Red lines show the trace of the base Syu surface interpreted as unconformity by Le Deit et al. (2013), which has >1 km of relief. ★=sedimentary-mound summit. (b, c) Colors show paleo-topography of Gale's mound, interpolated using (b) inverse-distance weighting and (c) quadratic polynomial interpolation. Black contours show isochores for late-deposited material. ▲,△=high points of unconformity surfaces (filled for IDW-interpolated, open for quadratic-interpolated). ×=locations of maximum thickness for upper units. (d, e, f): As (a-c), but for base Bu surface interpreted as unconformity from Le Deit et al. (2013) (this surface is best-fit by a saddle: see Table 3).

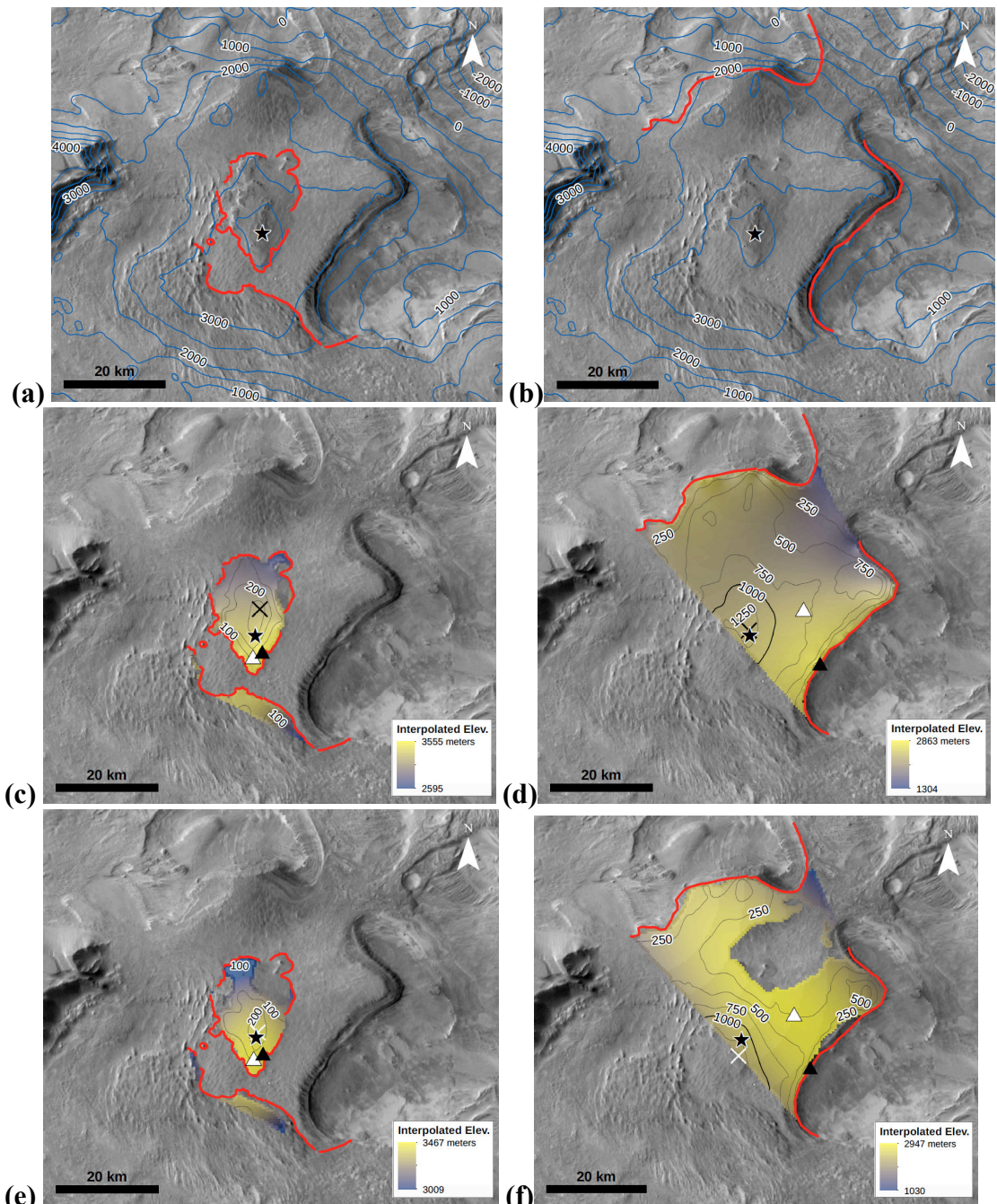

**Fig. 12.** Paleo-domes within Mars mounds. (a) Topography (blue contours) for Ceti Mensa (Candor Chasma is ~120 km wide). Red lines show the trace of base Caprock surface interpreted as unconformity by Lucchitta (2015), which has >1 km of relief. ★=sedimentary-mound summit. (b, c) Colors show paleo-topography of Ceti Mensa, interpolated using (b) inverse-distance weighting and (c) quadratic polynomial interpolation. Black contours show isochores for late-deposited material. ▲,△=high points of unconformity surfaces (filled for IDW-interpolated, open for quadratic-interpolated). ×=locations of maximum thickness for upper units. (d, e, f): As (a-c), but for base Rimrock surface interpreted as unconformity by Lucchitta (2015). Background is THEMIS VIS mosaic.

| Location | Mound | Stratigraphic surface (interpreted as unconformity) | RMS error on planar fit | RMS error on quadratic fit | Quadratic fit | Relief on stratigraphic surface (interpreted as unconformity) | Maximum thickness above unconformity (IDW interpolation)[1] | Maximum thickness above unconformity (quadratic interpolation) |
|---|---|---|---|---|---|---|---|---|
| W Candor | Ceti Mensa | base caprock[2] (Lucchitta 2015) | 500 m | 204 m | Saddle, high axis oriented 95° CW from N, saddle located near 76.3° W 5.8S. | 3.2 km | 1.4 km | 0.7 km |
| W Candor | Ceti Mensa | base caprock (Lucchitta 2015), high confidence region | 301 m | 77 m | Dome, ellipticity 1.4, centered at 76.35°W 5.87°S, long axis 6° CW from N. | 1.8 km | 0.4 km | 0.3 km |
| W Candor | Ceti Mensa | base Rimrock[2] (Lucchitta 2015) | 475 m | 168 m | Dome, ellipticity 3.2, centered 76.16°W 5.71°S, long axis 32° CCW from N. | 2.7 km | 1.4 km | 1.5 km |
| Gale | Mt. Sharp / Aeolis Mons | base Bu (Le Deit et al. 2013) | 154 m | 86 m | Saddle, high axis oriented 13° CCW from N, saddle located near 137.8°E 4.7°S. | 2.4 km | 1.5 km | 0.9 km |
| Gale | Mt. Sharp / Aeolis Mons | base (Syu2 + Cyu) (Le Deit et al. 2013) | 220 m | 132 m | Dome, ellipticity 1.6, center 137.86°E 5.08°S, long axis 96° CW from N. | 3.3 km | 2.7 km | 2.1 km |
| Gale | Mt. Sharp / Aeolis Mons | base 'Upper mound (Um) formation' (Thompson et al. 2011) | 826 m | 426 m | Dome, ellipticity 1.3, center 137.80°E 5.25°S, long axis 41°CW from N. | 4.0 km | 2.8 km | 2.1 km |
| Gale | Mt. Sharp / Aeolis Mons | base 'Upper unit' (Anderson & Bell 2010) | 219 m | 114 m | Dome, ellipticity 1.7, center 137.88°E 4.77°S, long axis oriented 13°CW from N. | 1.4 km | 1.6 km | 0.6 km. |
| Ophir* | Ophir Mensa * | "Marker horizon" (Wendt et al. 2011 / Peralta et al. 2015) | 850 m* | 339 m* | Dome, center 73.54°W 3.99°S, ellipticity 1.8, long axis 86° CW from N. * | 5.2 km* | 1.2 km* | 1.1 km* |

**Table 3.** Table of mound-spanning stratigraphic surfaces interpreted as unconformities. *Notes:* 1. 100 nearest-neighbouring points (i.e., vertices on unconformity trace), 100km search radius, quadratic weighting. 2. The 'ildu' of Fortezzo et al. (2016) corresponds to base rimrock in places, and to base caprock in other places. However, fitting a quadratic surface to the Fortezzo et al. (2016) 'ildu' would also produce a dome. *Doubtful. The "marker horizon" trace follows the top of a cliff near 74°E 4°S, in an area without stratigraphic cues; all measurements associated with this trace are lower confidence.

## 4.2. Draped landslides.

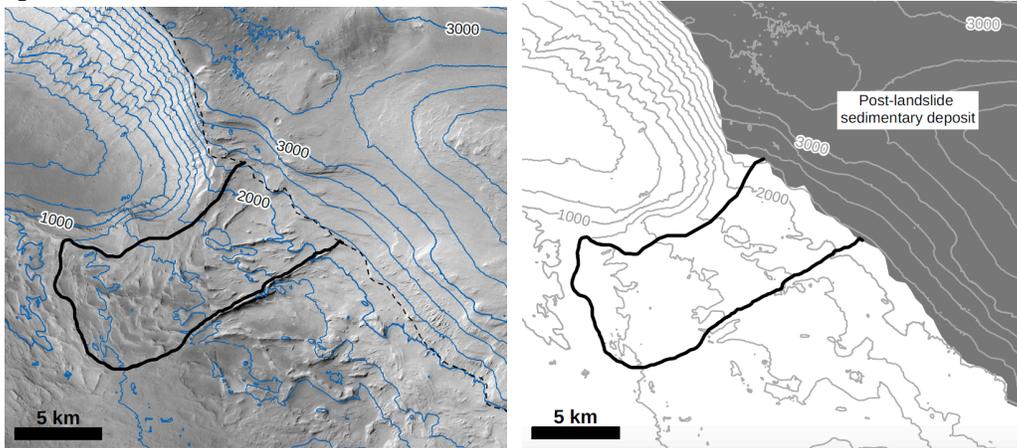

**Fig. 13.** Draped landslides in E. Candor. **(a)** *Left panel:* landslide on S side of Juventae Mensa (65.9W 8.1S). Range of elevations ~3 km. *Right panel:* Sketch interpretation (200m topographic contours). Crosshatching denotes post-slide sedimentary rock. All parts of figure use CTX DTMs. Location indicated in Figure 8a.

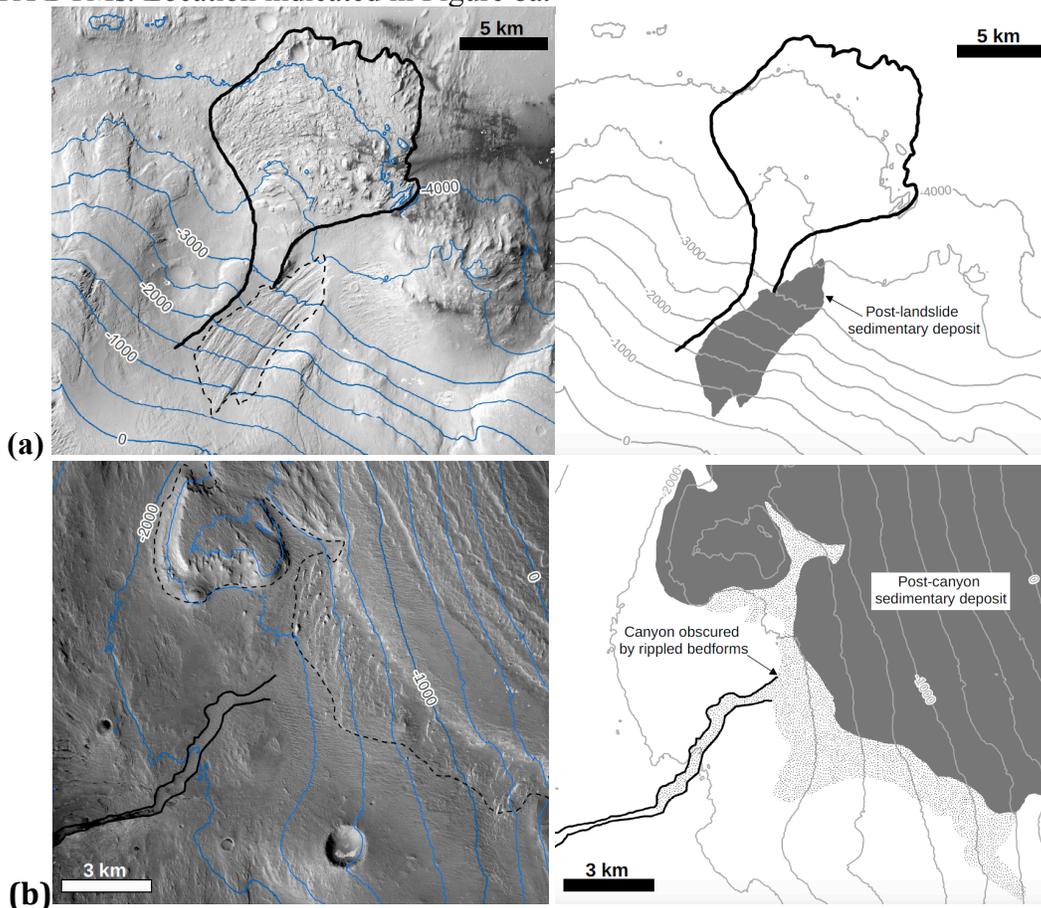

**Fig. 14.** Draped landslides and draped canyons in Gale. **(a)** *Left panel:* landslide on N side of Gale's mound (137.9E 4.8S). Range of elevations ~4 km. *Right panel:* Sketch interpretation (200 m topographic contours). **(b)** *Left panel:* Draped landslide on W side of Gale's mound (137.5E 5.1S). Range of elevations ~2 km. *Right panel:* Sketch interpretation (200 m topographic contours). Diamonds denote post-canyon sedimentary rock. Stippling denotes mobile cover. All parts of figure use CTX DTMs. Locations indicated in Figure 8f.

Gravity-slide deposits, when interstratified with sedimentary rocks, point away from paleo-highs on unconformity surfaces (Sharp 1940). Landslides encircling Ceti, Coprates, and Juventae Mensae, Gale's mound, and possibly Melas Mensa, flowed away from mound crests and are overlain by sedimentary rocks, especially at locally high elevations (Lucchitta 1990, Neuffer & Schultz 2006, Okubo 2014) (Figs. 13-14). (A moatward-draining canyon at Gale's mound is also draped by sedimentary rocks; Fig. 12). Therefore, sedimentary rock emplacement on topographic highs continued after moats were defined. Therefore, draped landslides record are diagnostic for paleo-moats. Draped landslides exclude a scenario in which paleo-domes result from rapid differential compaction of initially-horizontal layers, because that scenario does not permit dome-shaped syndepositional paleotopography. To the contrary, draped landslides suggest that the paleo-dome unconformities had an erosional origin.

## 5. Assessment of mound emplacement hypotheses, emphasizing Valles Marineris.

The VM mound-emplacement hypotheses that are most frequently discussed are the compensational-stacking (playa/lake or fluviodeltaic infill) and anticompensational stacking (e.g. slope winds) models (Section 1).

Other VM ILD formation models (Nedell et al. 1987, Lucchitta 1992) include nunutaks (Gourronc et al. 2014), tuyas (Chapman & Tanaka 2001), spring mounds (Rossi et al. 2008), salt-sheet outliers (Montgomery et al. 2009), salt tectonics (Jackson et al. 1991, 2011; Baioni 2013), and carbonate deposits (McKay & Nedell 1988) (Fig. 15). The tuya and carbonate mound hypotheses fail to match post-2004 spectroscopic data. We cannot logically exclude a scenario in which the VM mounds are volcaniclastic/ash/pyroclastic deposits emplaced on the flanks of a dyke or a central volcano. However, this possibility is disfavored by (i) the tendency of fissure eruptions to evolve to pipe-eruptions geologically quickly (Wylie et al. 1999), in contrast with the shapes of the VM mounds; (ii) the regular layering of the rhythmite, suggesting quasi-periodic deposition as opposed to the power-law behavior exhibited by volcanic eruptions (Lewis et al. 2008, Pyle 1998). The salt-sheet outliers hypothesis invokes a laterally continuous salt layer (extending under the plateaus encircling VM). This hypothesized layer is hard to reconcile with VM wallrock observations that do not show salt layers, or that show salt layers which drape onto wallrock. The nunataks hypothesis invokes wet-based glaciers for which there is little uncontested evidence. The spring mounds hypothesis has difficulty explaining the great lateral continuity of observed layers.

We cannot exclude the possibility that the VM mounds are giant salt domes. However, salt movement (Jackson et al. 1991, Jackson et al. 2011) after moat formation would be sideways, not upwards (as a salt glacier). Salt diapirism before moat erosion would not lead to systematic outward dips in outcrop. Where diapirism is inferred on Mars, it has a horizontal length scale that is comparable to the thickness of the sedimentary layer, and so much less than the $\sim 10^2$ km length of the VM mounds (Bernhardt et al. 2016). Faulting can and does tilt layers (Lewis & Aharonson 2014), but syndepositional basement uplifts beneath (and only beneath) mounds are unlikely. In particular, we disagree with the syndepositional-tilting proposal of Fueten et al. (2008) because the upper materials – the "caprock" and "rimrock" of Lucchitta (2015) – lack obvious major faults. Predepositional tectonic uplifts might nucleate draping deposition, but draping deposition on highs is an example of anticompensational stacking, not an alternative. Landslides are common, but are easy to identify and are excluded from our layer orientation measurements (Fig. 11).

Differential compaction of sedimentary layers over basement relief has been proposed to reconcile deposition of flat-lying strata with observed layer-orientations (Grotzinger et al. 2015). This is inconsistent with VM data. In VM, the basement surface of canyon floors that are not covered by thick sedimentary deposits is observed to be flat (e.g., E Coprates Chasma, Ganges Chasma, Noctis Labyrinthis). The implication that the central Valles Marineris canyons formed via near-vertical tectonic subsidence is strongly supported by independent tectonics data and modeling (Andrews-Hanna 2012b). If the basement of the VM canyon floors is flat before sedimentary loading, then after flexural subsidence the basement will dip inwards. This inward dip should set the sign of differential compaction tilts for initially flat-lying strata with uniform grain-size. If grain-size is not uniform, then differential compaction can tilt layers away from coarse-grained deposits. However, fluviolacustrine processes will preferentially deposit coarse grains near the margins of the canyon, again leading to inwards dips. Therefore, if differential compaction caused layer-dips in VM, we should not see outward dips on both sides of a mound. However, we do observe outward dips on both sides of mounds (Fig. 6, Fig. 8), contradicting the hypothesis of differential compaction for layer-orientations in VM mounds.

A hybrid hypothesis could reconcile VM layer orientations with initially horizontal deposition. In this hypothesis, early-deposited sediments were first pre-compacted by thick overburden, and then eroded into wedge-shaped outliers. A later generation of sediments was differentially compacted over these wedge-shaped outliers, leading to the observed outward-dipping layers (Fig. 16). This hypothesis is a hybrid because terrain-influenced winds are required to erode the pre-compacted sediments into a correctly-shaped wedge prior to further sediment deposition. Even if a buried wedge of ancient sediments exists and was pre-compacted sufficiently to act as a rigid floor for later differential compaction, compaction is at best marginally sufficient to explain the large amplitude of observed dips (Gabasova & Kite 2016). Furthermore, if hypothetical wedge-shaped remnant deposits exist, then they exist mainly in subcrop, because inspection of HiRISE images does not show the large-scale within-mound onlap predicted by this scenario.

Other mechanisms that rotate layers outwards during mound construction are quantitatively insufficient to explain the data, require fine-tuning, or both. For example, flexural tilting due to late-stage volcanism is <0.1° (Isherwood et al. 2013); outward tilting by the flexural response to erosional unroofing is <0.2° (Davis 2007) and can only partly recover inward tilting during sedimentary rock loading; and post-Noachian crustal-flow is minor (Karimi et al. 2016).

These considerations favor the interpretation that the dip directions are primary, i.e. that the mounds grew as mounds, and that present-day mound crests are close to the crests of the growing mounds (Anderson & Bell 2010) (Figs. 2-3). In combination with the paleo-dome and draped-landslide evidence, the layer-dips suggest anticompensational stacking.

One mechanism that predicts anticompensational stacking is slope-wind intensification of erosion on steep topographic slopes (Kite et al. 2013a, Day & Kocurek 2016). In this model, terrain-induced winds inhibit sedimentary rock emplacement on crater/canyon walls, creating paleo-moats. These paleo-moats serve as the basal surface for subsequent deposition. Slope-wind controlled sedimentary-basin build-up combines processes that individually have a well-understood terrestrial analog but which rarely occur in combination on Earth. For example, katabatic winds drain the Antarctic plateau (Parish & Bromwich 1991); deep incision into rock by wind erosion has been reported from the Atacama (Perkins et al. 2015); and the Qaidam basin is being exhumed by wind erosion (Heermance et al. 2013).

Slope-wind dynamics are not the only means of producing anticompensational-stacking kinematics: snow destabilization by föhn winds (Brothers et al. 2013), reduced saltation-transport to higher elevations due to lower pressure, and greater availability of abrasive sand at lower elevations, could all cause preferential net erosion of sediments at lower elevations – and thus favor anticompensational stacking. Sediment can be delivered by suspension transport ("airfall") and also by saltation transport. Saltation transport to paleohighs need not be prevented by moats; sand dunes flow uphill in modern VM and on polar mounds (Chojnacki et al. 2010, Conway et al. 2012; another example is visible in ESP_029504_1745).

These slope-dependent models have the common advantage that they all predict that outward-directed dips should be ubiquitous, provided that craters/canyons have long, steep walls (Figure DR2 in Kite et al. 2013a). This matches our observations – outward dips are very common (section 3).

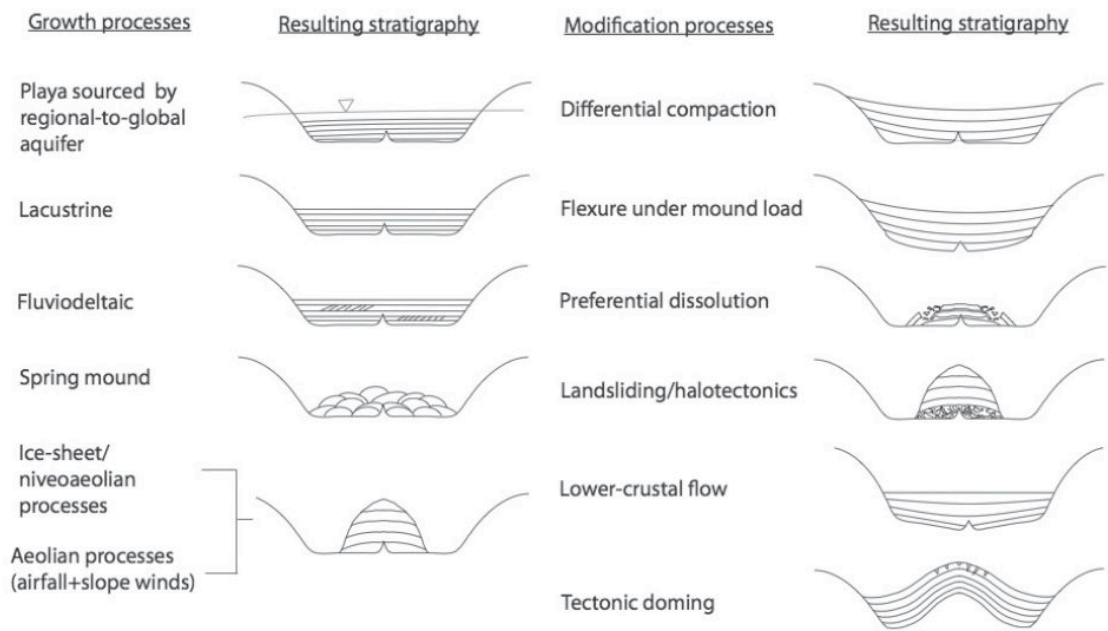

**Fig. 15.** (Modified from Kite et al. 2013a). Comparison of the layer orientations predicted by different mound growth hypotheses, for an idealized cross section of a mound-bearing crater. Inverted triangle marks past water table.

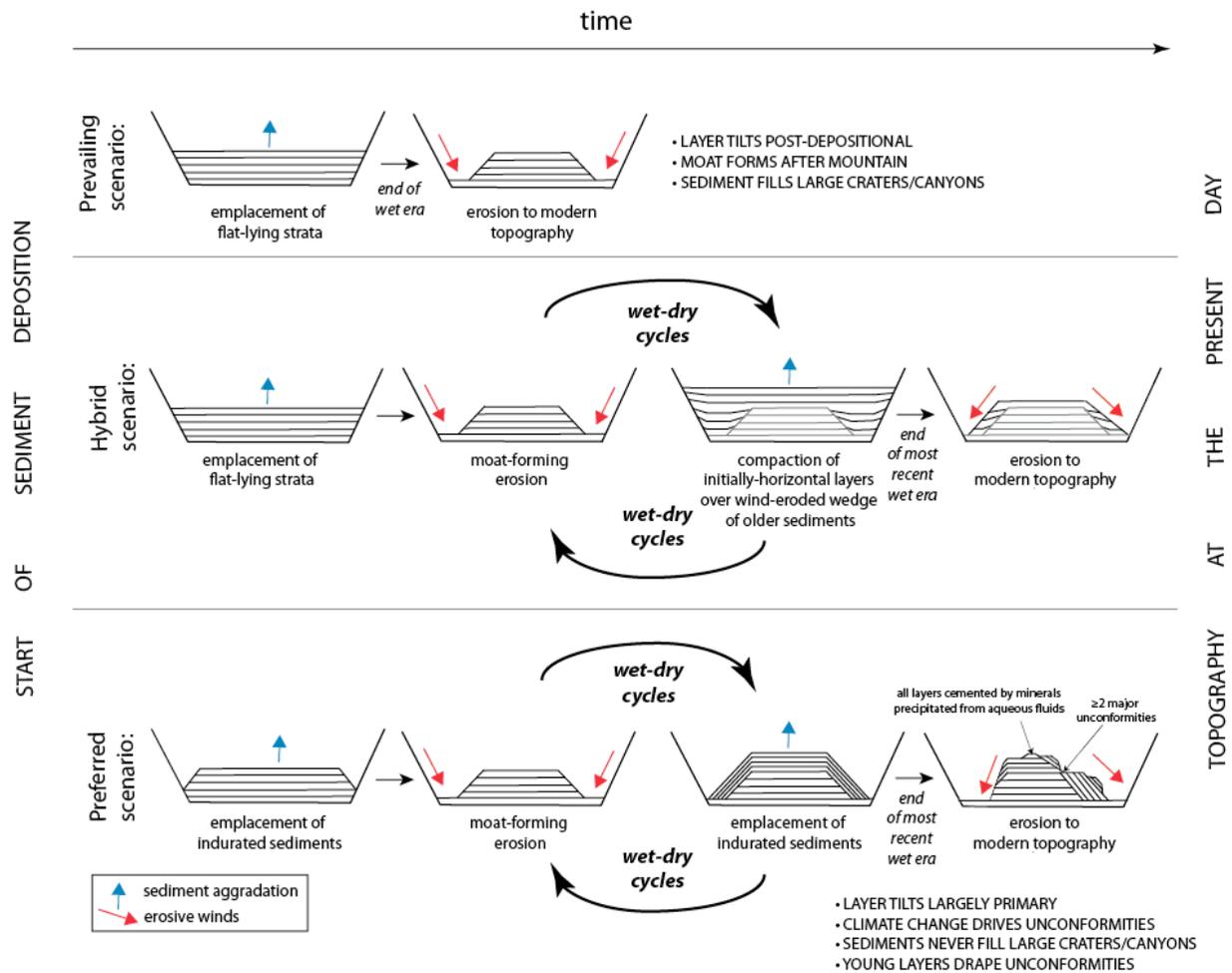

**Fig. 16.** Summaries of 'prevailing view,' 'hybrid scenario,' and 'preferred view' of mound formation.

## 6. Model: anticompensational stacking and climate change.

Anticompensational stacking implies that layers steepen over time. Steepening could occur via layer truncation at unconformities, mound-scale layer pinch-out, or both. We did not find evidence for mound-scale layer pinch-out. Instead, we found layer truncation at a small number of large unconformities. We interpret these unconformities as paleomoat bounding surfaces (Figs. 8-9, Table 3) (Okubo et al. 2014). We infer that anticompensational stacking arises from long depositional intervals separated by major erosive intervals (Fig. 16) – a drape-and-scrape cycle.

Wind erosion can form paleomoats. Wind-induced saltation-abrasion is widely accepted to erode mound material in the present epoch (Grotzinger 2014), to have formed present-day moats (Kite et al. 2013a, Day & Kocurek 2016), and to have had greater erosive power in Mars' past. To parameterize moat and paleomoat formation, we used Mars Regional Atmospheric Modeling System (MRAMS) simulations (Rafkin et al. 2001), a realistic day-night cycle, and idealized mound-and-moat topography (Appendix B). MRAMS results indicate that slope effects are crucial to moat formation at low atmospheric pressure, with wind stresses ~5× greater on steep slopes relative to flat floors within craters/canyons (Fig. B3). Higher wind stresses are likely correlated with faster long-term wind-erosion, because wind stress is observed to exert strong control on aeolian sediment transport rates on Mars (Ayoub et al. 2014), and because aeolian sediment transport is required to provide abrasive particles for sandblasting and/or to remove debris.

Mars' obliquity ($\varphi$) varies quasi-periodically at $10^5$-yr timescales, but chaotically at longer timescales, ranging from 0°-70° (Laskar et al. 2004), with significant effects on climate. At low $\varphi$, models indicate that water is less available at the low latitude of VM and Gale (e.g. Jakosky & Carr 1985, Mischna et al. 2003, Madeleine et al. 2009, Andrews-Hanna & Lewis 2011, Mischna et al. 2013, Wordsworth et al. 2013, Kite et al. 2013b). Without water for cementation, sediment does not get preserved in the sedimentary rock record. Sedimentary rock formation is also disfavored by surface condensation of atmospheric $CO_2$ at low-$\varphi$ (Forget et al. 2013, Soto et al. 2015); atmospheric collapse suppresses aeolian-sediment supply and surface liquid water. At high $\varphi$, by contrast, water is progressively driven to lower latitudes as polar regions receive greater insolation (e.g. Mischna et al. 2013). Additionally, sediment deposition rates may be enhanced by globe-spanning storms expected at high $\varphi$ (Haberle et al. 2003, Armstrong & Leovy 2005, Newman et al. 2005). These considerations suggest an important role for $\varphi$ in modulating sedimentary rock build-up (Lewis et al. 2008). In the words of Metz et al. (2009), "Obliquity-driven climate […] may be a more significant factor in the development of the stratigraphic record of Mars as compared to Earth."

To model mound build-up including chaotic $\varphi$ forcing and paleomoat formation, we carried out >100 simulations of Mars $\varphi$ history. Each simulation combines 3-Gyr long 8-planet `mercury6` (Chambers 1999) simulations and an obliquity model (Armstrong et al. 2004, 2014). For each simulation, we assume sedimentary rock accumulation (assumed, for simplicity, to occur at a spatially uniform rate) competes with terrain-influenced erosion at VM and Gale when Mars' obliquity ($\varphi$) > 40°, but that erosion alone operates when $\varphi$ < 40°. The critical obliquity value is somewhat arbitrary, although all low-atmospheric-pressure models predict a nonlinear increase in the abundance of surface water ice at the latitude of VM and Gale at $\varphi > (40^{+5}_{-8})°$. We do not model the between-basin variation in availability of liquid water needed for cementation; previous work shows (Andrews-Hanna & Lewis 2011, Kite et al. 2013b) that between-basin variability can match the scenario presented here. We also do not model the $<10^5$ yr-timescale cycles that are responsible for the development of the layers whose orientation we measure, because these cycles occur at much shorter timescales than the overall mound construction modeled here (Lewis & Aharonson 2014). Possible causes of layering are discussed in (e.g.) Kite et al. (2013b) and Andrews-Hanna & Lewis (2011). These simplifications ensure that the details of the model do not obscure the processes modeled by `SOURED` (slope-wind control of within-basin spatial variations, and nonlinear obliquity control of mound-spanning unconformities variations). To combine obliquity forcing and wind-terrain feedback, we use a 2D (horizontal-vertical) landscape evolution model. Sediment is supplied from distant sources, and eroded material is removed to distant sinks. Consistent with CTX-scale morphology, thermal inertia, and the paucity of craters on sedimentary mounds (Malin et al. 2007), we assume that sedimentary rocks are much more erodible than igneous "basement". We adjust accumulation-rate so that modeled mounds are ~3 km tall. The model produces mounds of the correct height with a mound-sediment deposition rate $D \approx 25$ μm/yr, a rate that is independently suggested by the thicknesses of orbitally-paced layers (Lewis & Aharonson 2014). Model maximum deposition-rates are similar to maximum erosion-rates. Higher wind-erosion rates 3 Gya are consistent with $O(1)$ μm/year modern-era wind-erosion rates (Grindrod & Warner 2014, Farley et al. 2014) if the supply of abrading particles is not limiting, wet-era atmospheric pressure was ~60 mbar (Catling 2009, Brain & Jakosky 1998), and sandblasting rate increases faster than linearly with atmospheric pressure (Kok et al. 2012).

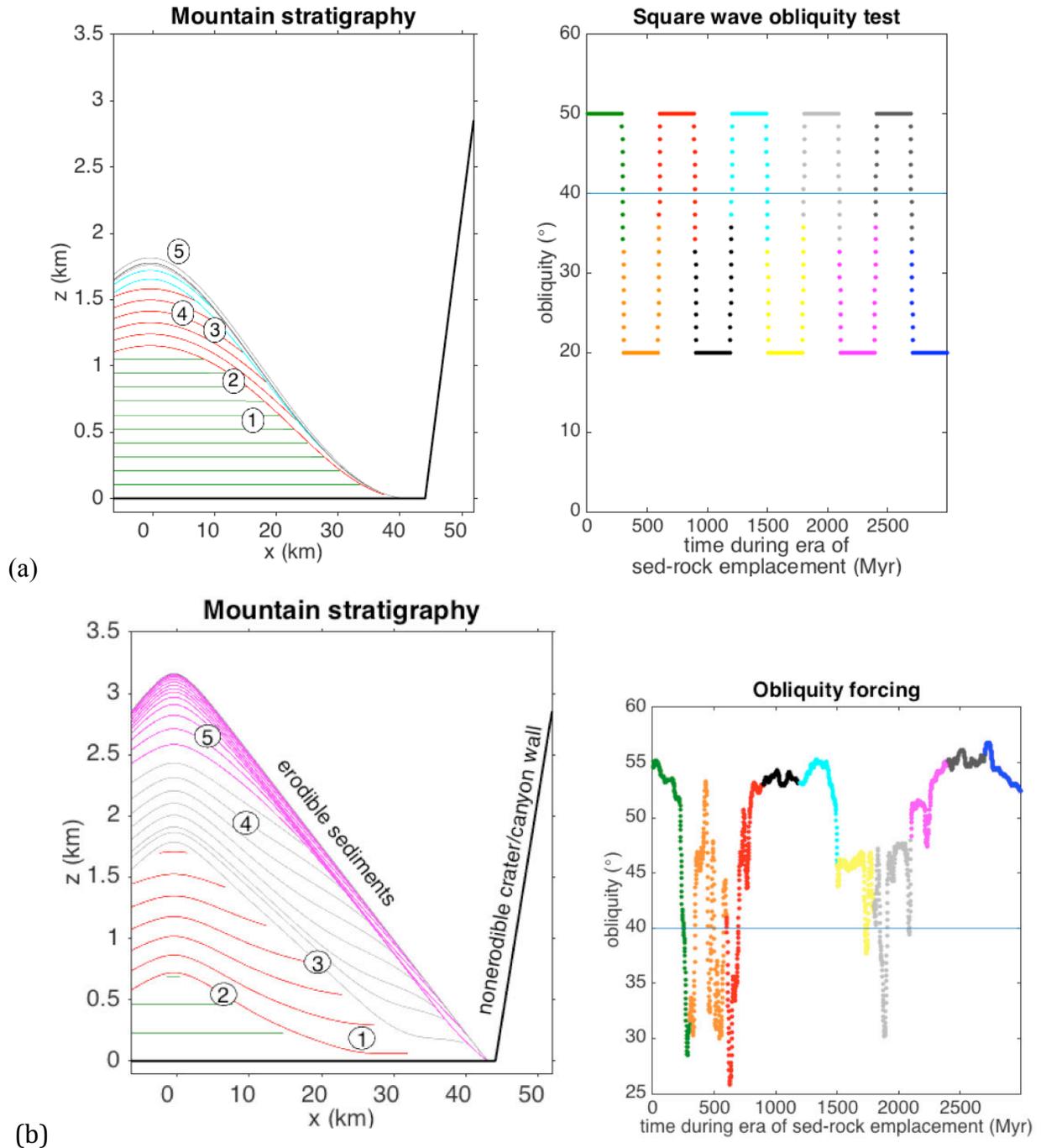

(a)

(b)

**Fig. 17.** Model of how Mars mound stratigraphy might encode chaotic climate change. (a) Square-wave demonstration of how obliquity forcing and slope-winds combine to explain the basin-scale stratigraphy of the largest sedimentary rock mounds on Mars. *Right panel:* Alternations every 300 Ma between high mean obliquity (deposition) and low mean obliquity (no deposition, erosion only). Critical obliquity shown by horizontal blue line. *Left panel:* Sedimentary stratigraphy shown by colored line. Black line is nonerodible container. Lines are drawn at 20 Myr intervals. Colors change every 300 Myr. The numbered properties of the model

output are consistent with the data: ① First-deposited stratigraphic package has layers that gently dip away from the mound center (Fig. 6c). ② Unconformities slope away from mound center, defining paleo-domes (Fig. 9-10). ③ Unconformities steepen up-mound. ④ Dips steepen up-mound (Fig. 7). ⑤ Unconformity-bounded stratigraphic packages thin moving up-mound (Table 3). (b) *Right panel:* One possible history of orbital forcing. Horizontal line shows critical obliquity above which sedimentary rock emplacement is permitted. *Left panel:* Stratigraphy of mound formation for this orbital forcing (late-stage erosion is not shown). Colored lines show stratigraphy. Black line is nonerodible container.

Model output (Figs. 17-18) shows that aeolian sedimentary rock emplacement forced by chaotic $\varphi$ change, and including the wind-terrain feedback effect, can produce free-standing mounds within a crater/canyon (Kite et al. 2013a). The basic implications of obliquity forcing for slope winds are illustrated in Fig. 17. Fig 17a uses square-wave deposition forcing, and Fig. 17b uses an example realistic forcing. In reality, obliquity is chaotic; many simulations are needed to bracket the range of possible behavior (e.g. Fig. 18). As expected, the dominant behavior is anti-compensational stacking.

A key attribute of modeled sedimentary deposits (Figs. 17-18) is that both layers and internal unconformities dip away from mound crests, consistent with data (Fig. 6-8). Though the mound topography and pattern of outward dip directions observed within Mars' sedimentary rock mounds are the most prominent features explained by this mechanism, the predicted stratigraphy simultaneously matches a range of observed physical attributes. These include the average dip magnitudes (which cluster at the mound height:width ratio), the thinning-upwards of unconformity-bounded stratigraphic packages (Malin & Edgett 2000), the layer thicknesses, and the outward dip of unconformities (Figs. 2-3) (Banham et al. 2016). The results explain why layer orientations frequently conform to modern topographic slope (Fueten et al. 2008). Because deposition occurs by progressive draping on pre-existing mound topography, only strongly nonuniform erosion (e.g. the canyons incised into Gale's mound) can create slopes that greatly differ from layer orientations. In Mars' mounds, layers are predicted to steepen upward in the stratigraphy, as subsequent layers jacket a more-gently-dipping mound core; the opposite of the geometry encountered in mountains on Earth. Modeled dips tend to steepen up-mound. The dips of *exposed* layers can either steepen up-mound or remain constant, depending on the depth of late-stage erosion. Because chaotic shifts in mean obliquity are infrequent (Lissauer et al. 2012, Li & Batygin 2014), the 1-2 large unconformities observed in some mounds suggest a (discontinuous) span of liquid water $\gg$100 Myr long. This is consistent with the ~100 Myr lower bound estimated by rhythmic layering using only the thickness of the preserved sedimentary rock, and not accounting for unconformities (Lewis & Aharonson 2014).

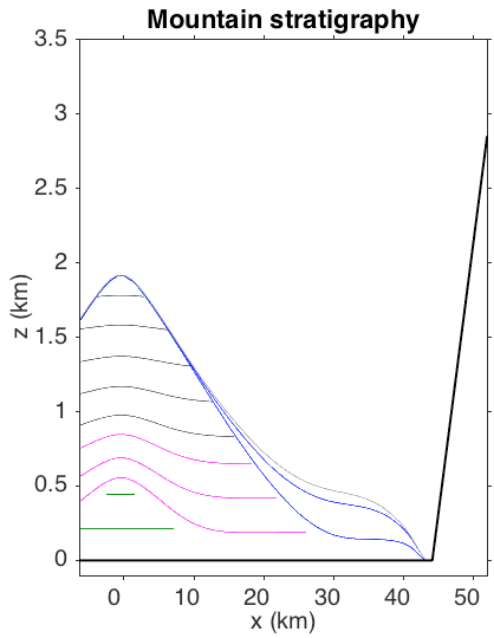

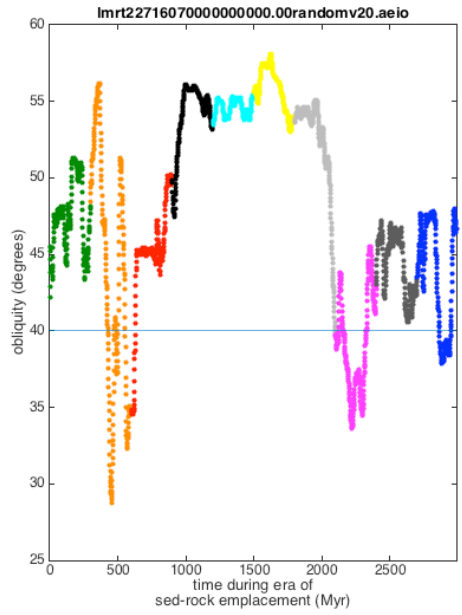

a)

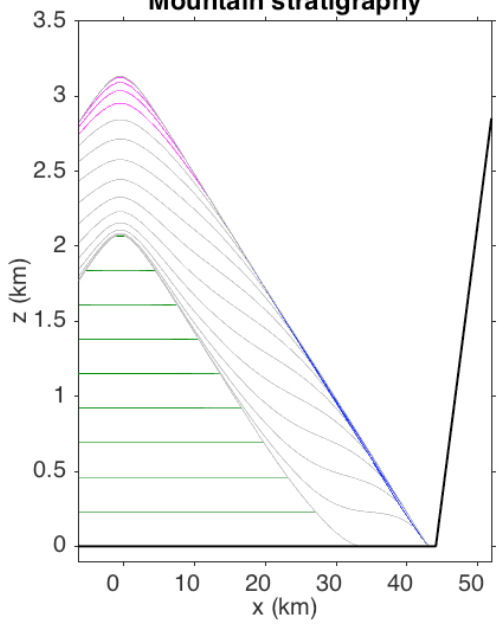

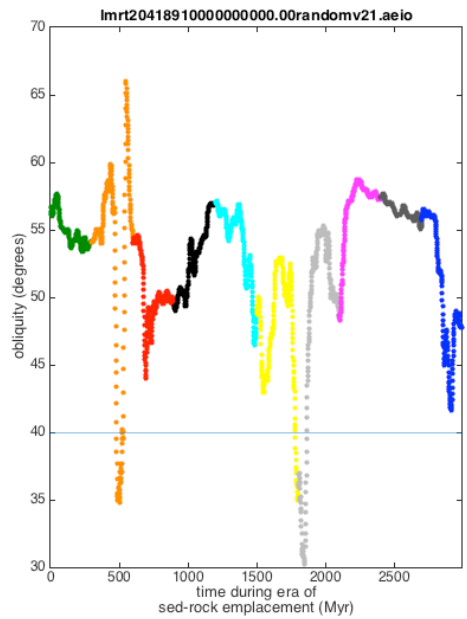

b)

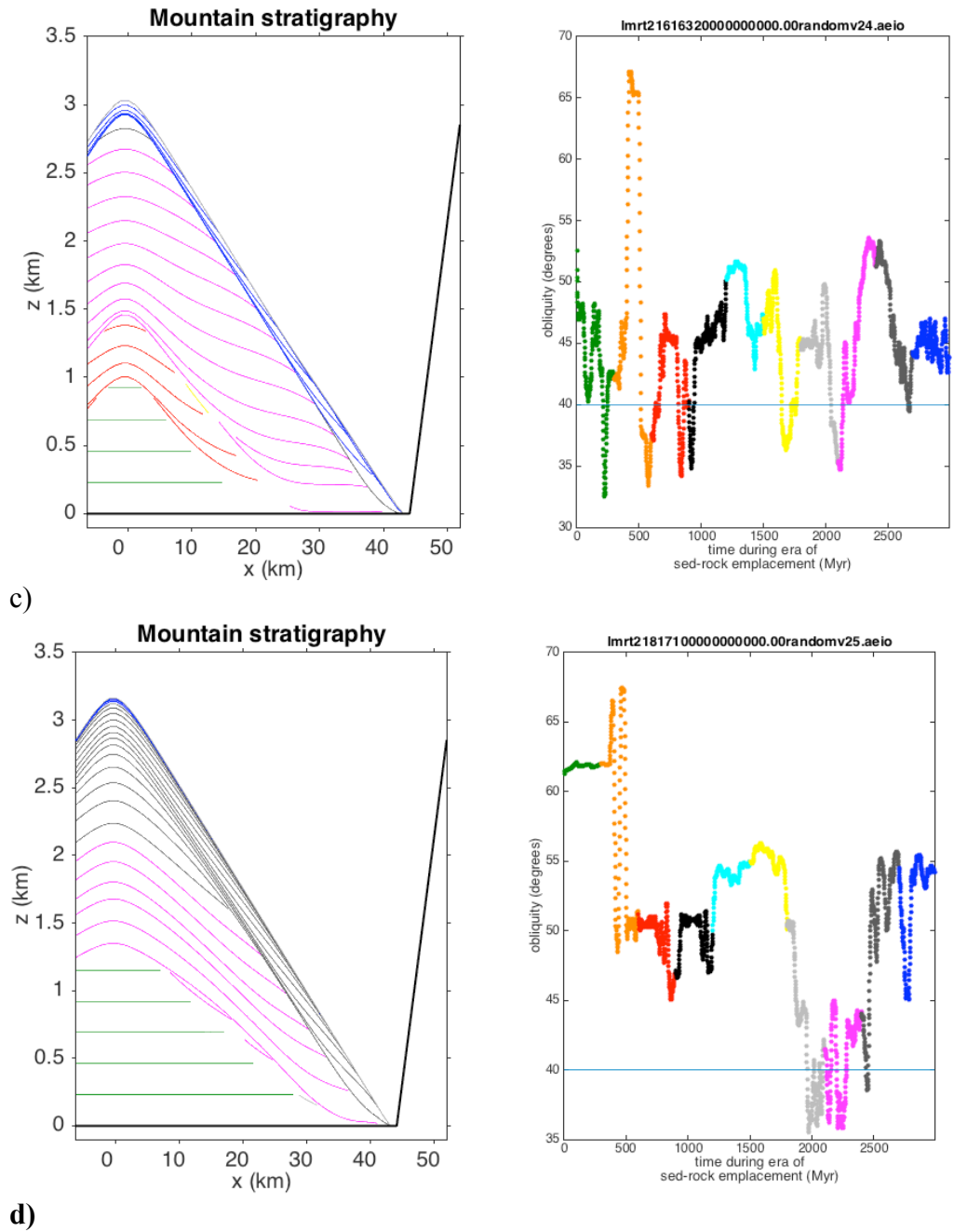

**c)**

**d)**

**Fig. 18.** Additional examples of mound stratigraphies with their corresponding obliquity forcing, chosen to illustrate a range of interesting behavior. Layers drawn every 20 Myr of simulated time, color change every 300 Myr. Notice the within-moat depositional package in (a), and the "scabbed" depositional packages on the mound flank in (c) and (d).

# 7. Discussion.

## 7.1. Limitations of data interpretation.

Anticompensational stacking can explain most of the layer orientations of most >1 km-thick sulfate-dominated stratigraphies within deep and steep-sided craters/canyons. (The residuals might be due to gravity-driven slumping, or to deposition onto a paleo-surface that had been wind-eroded into a non-axisymmetric shape - e.g., the present topography of the mound in Nicholson crater). However, many Mars mound stratigraphies do not fall into this category. For example, sedimentary mounds in Terby crater (Ansan et al. 2011) show a complicated 3D stratal architecture that cannot be reproduced by the 2D slope-winds model used here (Wilson et al. 2007, Ansan et al. 2011). Mounds within craters in W. Arabia Terra have been argued to be outliers of a formerly more extensive deposit on the basis of geographic continuity (Bennett & Bell 2016). Geographic continuity makes predictions regarding the ice mounds encircling Mars' North Polar Layered Deposits that are known to be incorrect (Conway et al. 2012, 2013; Brothers et al. 2013; Brothers & Holt 2016); therefore, geographic continuity is inconclusive. Layer dip data are unavailable for these Arabia mounds.

A second key limitation of the anticompensational-stacking interpretation is that it does not work for small deposits. For example, small catenae contain layered deposits that dip inward (e.g. Weitz & Bishop 2016), and small craters in Arabia show inward dipping layers in anaglyph (e.g. HiRISE PSP_001981_1825/PSP_0012258_1825). This proves that for small container size ($\ll$ 100 km), anticompensational stacking is not effective. In turn, this suggests a critical length/depth scale above which slope winds are most effective (§2.5). This means that our MRAMS mesoscale results need not contradict the Day et al. (2016) Large Eddy Simulation study (which emphasizes the role of unidirectional winds), but could simply refer to a different (> 100 km) scale of Mars crater/canyon.

## 7.2. Assumptions and limitations of model.

The biggest uncertainty in our landscape evolution model is sediment availability. Sediment is assumed to be available for sedimentary rock emplacement during depositional intervals, and sand is also assumed to be available for sandblasting. This assumption of "sufficient" sand/dust/ash in Mars' past is motivated by modern data. Today, sand is present almost everywhere, but (except in a few places) is probably not pervasive and persistent enough to armor steeply-sloping bedrock over geologic time (Hayward et al. 2014). Present-day gross dust accumulation rates are not much less than inferred ancient sediment accumulation rates (Kinch et al. 2007, Lewis & Aharonson 2014). The rate of production of fine-grained material would be greater in the past because the rates of volcanism, impacts, physical erosion, and chemical weathering were all greater in the past (e.g. Golombek et al. 2006, Levy et al. 2016, Carter et al. 2013). This motivates the assumption that over long timescales, sediment is not limiting. On shorter timescales that are not resolved by the landscape evolution model, peaks in both erosion and deposition will probably be tied to the passage of supplies of abundant sand (so sediment starvation might control mound build-up at <1 km stratigraphic scales). Even if sediment is available, it will not stay in one place for Gyr unless liquid water is available to indurate it. Because our data indicate that mound build-up continued after a topographic moat was defined, regional groundwater flow is implausible as a water source for those upper layers and so the water needed for cementation must be from a top-down water source such as rain or snowmelt (Clow 1987, Niles et al. 2009, Kite et al. 2013b, Fairén et al. 2014).

The most important assumption in the wind erosion model is that output from 6 mbar simulations is relevant to the times when most erosion (and sedimentation) occurred, when the atmospheric pressure was likely higher (e.g. Catling et al. 2009). Since the absolute erosion rate is nondimensionalized in our model, only the pattern of wind erosion matters. Strong slope winds are expected on long steep slopes provided that the atmosphere is thin enough to permit large day-night swings in temperature (Zardi & Whiteman 2013). Thus we expect that terrain strongly influenced wind-erosion patterns in Mars' past.

Patches of layered deposits veneer the slopes of some of the VM canyons (e.g. Fueten et al. 2010, 2011). Pasted-on wall-slope deposits can form in our model, but tend to be removed by late-stage erosion. The observed persistence of these outliers highlights the limitations of our 2D modeling approach. To investigate these outliers would require a fully coupled 3D model of landscape-wind coevolution.

Although anticompensational stacking is the dominant behavior in our model, we did find cases where the slope winds model places lenses of sedimentary rock low down on the mound or in the moat (e.g. Fig. 18a). These packages correspond to late-stage materials that are on close-to-modern topography. Possible real-world examples are (1) young materials in the moat SW of Ceti Mensa (Okubo 2010), (2) the light-toned yardang-forming unit towards which the Mars Science Laboratory rover is driving, (3) the Siccar Point group in Gale including the Stimson formation (Fraeman et al. 2016).

In our model there is no *secular* climate change. This is unrealistic; secular climate change clearly occurred on Mars (Jakosky & Phillips 2001). Our calculations assume that climate change driven by chaotic alternations in mean obliquity introduces a large-amplitude overprint on secular change, and we focus on those alternations.

### 7.3. Geological implications and tests.

Obliquity strongly influences the three limiting factors for sedimentary rock build-up on Mars: sediment supply, water supply, and erosion intensity. Orbitally-forced drape-and-scrape cycles produce a good match to observations (Fig. 17). However, alternatives to $\varphi$-modulated accumulation exist. Secular variations in sediment supply, induced for example by regionally-coordinated volcanism, could explain the unconformities. This could be tested by mapping longitudinal trends in unconformity patterns. Alternatively, volcanism might globally coordinate wet episodes via greenhouse forcing. However, volcanic greenhouse gases are either too long-lived ($CO_2$) or too short-lived ($SO_2$) to easily explain the modulations (Kerber et al. 2015). Ice/dust cover might intermittently shield rocks from abrasion, but latitudinal shifts of cover materials are likely to be themselves $\varphi$-paced. If the great unconformities are obliquity paced, then the time gaps at unconformities should be >100 Ma. This can be tested via counts of embedded craters. Only one time gap at a Mars unconformity has been constrained so far (Kite et al. 2015), and the time gap is found to be >100 Ma, as predicted.

Latitudinal variations offer clues to mound origin. The biggest sedimentary mounds on Mars lie near the equator. These mounds have few obvious mound-spanning angular unconformities. By contrast, mounds poleward of ±25° (e.g. Galle, Terby) show numerous unconformities. This is expected for deposits forming at the margins of the latitudinal belt that permitted sedimentary rock formation (Kite et al. 2013b). The variation in mound height between canyons (the thickest deposits are in Northern VM, i.e. closer to Mars' Equator) could be due to a preference for sedimentary rock emplacement near the equator (Kite et al. 2013b). Alternatively, greater erosion in the canyons that now have thinner deposits might explain the latitudinal trend. Tests include measuring layer thicknesses (e.g. Lewis & Aharonson 2014, Cadieux & Kah 2015) and unconformity spacings.

Obliquity-modulated build-up can be tested by the *Curiosity* rover's climb through sulfate-bearing layers toward the major unconformity at Gale's mound identified by Malin & Edgett (2000). An origin via chaotic shifts in mean obliquity predicts that sedimentation episodes are long and few in number. $\varphi$-control predicts long time gaps at unconformities (Fig. 3), with gently dipping layers erosionally truncated and overlain by more-steeply-dipping layers draped over preexisting stratigraphy. Detection of gravels sourced from Gale's rim within strata high in Mt. Sharp / Aeolis Mons would disprove our model. Instead, aeolian (and reworked-aeolian) deposits should dominate. Evidence for paleo-erosion by wind should be common close to unconformities. Onlap at unconformities would support the hybrid hypothesis in Fig. 16, whereas draping at unconformities would support the preferred interpretation in Fig 16.

Mound formation processes are tightly linked to early-Mars runoff intermittency. Even small seasonal streams would suppress the sand migration that is required for saltation-driven erosion (Krapf 2003), and gravity-driven stream erosion would also suppress the anticompensational growth of mounds. Aeolian sediment supply can be reconciled with lakes in VM (Harrison & Chapman 2008) if climate permitted lakes for only a small percentage of years (Palucis et al. 2016, Buhler et al. 2014, Irwin et al. 2015). Wet-dry alternations during Mars' era of sedimentary-rock accumulation, including long dry periods, are predicted by our preferred scenario. Intermittent habitability is consistent with the persistence of surface olivine on Mars, and the detection in Gale mudstones of chemical markers for extreme aridity (Farley et al. 2016). Our data disfavor the long-standing hypothesis (McCauley 1978) that the VM outcrops are lake deposits, but are consistent with a lacustrine origin for outcrops below the base of the topographically defined mounds in VM and below the clay/sulfate transition at Gale (Grotzinger et al. 2015).

# 8. Conclusions.

We introduce new data and a new model for the evolution of eight major sedimentary mounds in Valles Marineris and Gale crater.

Data:
- Seven out of eight mounds investigated show layer orientations that dip systematically away from the mound centerline, with median dip 5° ($n = 308$).
- Layer-orientation data have a precision and accuracy that are sufficient for the purpose of constraining mound origin.
- Stratigraphic surfaces interpreted as major mound-spanning unconformities are well-fit by a dome-shape in 6 out of 8 cases.

Interpretation:
- When combined, the layer orientation data, draped landslides, and our interpretation of stratigraphic surfaces interpreted as unconformities, require primary deposition of layers on outward-tilted slopes for the topmost ~1 km of the mounds.

- Lower in the stratigraphy, the layer orientation data are consistent with either (i) primary deposition of layers on outward-tilted slopes (Kite et al. 2013a) or (ii) a hybrid hypothesis in which slope-wind erosion sculpts pre-compacted sediments that later act as wedge-shaped indentors for differential compaction of later-deposited sediments.

Model:
- We present a model that combines spatially-resolved forcing (from mesoscale meteorological simulations) and time-variable forcing (realistic orbital integrations) to make quantitative predictions for the evolution of the major sedimentary basins of Mars. The meteorological simulations confirm a strong trend of increasing wind stress with topographic slope within both craters and canyons.
- The model predicts that Mars mound stratigraphy emerges from a drape-and-scrape cycle.
- The model simultaneously matches the following mound attributes: (i) layers dip away from mound crests; (ii) internal unconformities have a dome shape; (iii) average dip magnitudes cluster at the mound height:width ratio; (iv) unconformity-bounded stratigraphic packages thin upwards; (v) layer orientations frequently conform to modern topographic slope.
- We propose that major mound-spanning unconformities within Mars mountains correspond to periods of low mean obliquity (Mischna et al. 2013, Kite et al. 2015). Because chaotic shifts in mean obliquity are infrequent, the 1-2 large unconformities observed in some mounds suggest a (discontinuous) span of liquid water ≫100 Myr long. In our model, the major mound-spanning unconformities (once correctly ordinated) can be used for planetwide correlation.
- On the Earth, first-order erosion-deposition alternations (Sloss 1963) are driven at a global scale by the Wilson cycle (via orogeny and eustasy). On Mars, climate changes driven by infrequent chaotic shifts in mean obliquity may play an analogous role in shaping the planet's sedimentary record.

**Acknowledgements.**
We thank Michael Lamb, Frank Fueten, Gene Schmidt, Chris Okubo, Leila Gabasova, John Grotzinger, Mackenzie Day, Jeff Barnes, Daniel Tyler, Claire Newman, Mark Richardson, John Armstrong, Bill Dietrich, Jasper Kok, Corey Fortezzo, Baerbel Lucchitta, Nathan Bridges, Brad Thomson, and David Rowley, for enlightening comments and discussions, and for sharing unpublished data. We thank four reviewers for timely, thorough, and thoroughly useful reviews. We thank the HiRISE team for maintaining the HiWish program, which provided multiple images that were valuable for this work. We thank UChicago's Research Computing Center. This work was financially supported by the U.S. taxpayer via NASA grant NNX15AH98G.
Layer-orientation data are available as a supplementary table. DTM scripts are written in Bash and are available for download from https://psd-repo.uchicago.edu/kite-lab/uchicago_asp_scripts/. DTMs and other data produced for this study may be obtained for unrestricted further use by contacting the lead author (kite@uchicago.edu).

# Appendix A. Stratigraphic Model.

## A.1. Overview and physical basis of stratigraphic model.

The central element of our `SOURED` (Fig. A1) is the forward model of landscape evolution and stratigraphy (section A.2), which incorporates time-varying sedimentary rock emplacement (assumed uniform within craters/canyons for simplicity) and spatially-varying feedback from slope-winds. Time-varying sedimentary rock emplacement is forced by 3-Gyr long integrations of the orbit and spin-pole orientation of Mars (section A.3). Spatially-varying feedback from slope-winds is forced by a mesoscale wind model (Appendix B). Essentially, `SOURED` = an upgraded version of `SWEET` + `MRAMS` + (`mercury6` + `oblique`) (Fig. A1).

Slope winds are important on Mars. These diurnally-reversing winds result from the combination of high relief and day-night temperature swings of up to 130K (e.g. Kass et al. 2003). Slope winds are particularly strong within the equatorial craters and canyons that host sedimentary rock mounds, where Coriolis effects are weak and relief can approach 10 km. The coupling between long, steep slopes and strong winds on Mars emerges from basic physical principles and is model-independent (Spiga et al. 2011, Kite et al. 2013a, Zardi & Whiteman 2013, Moreau et al. 2014, Tyler & Barnes 2015, Rafkin et al. 2016).

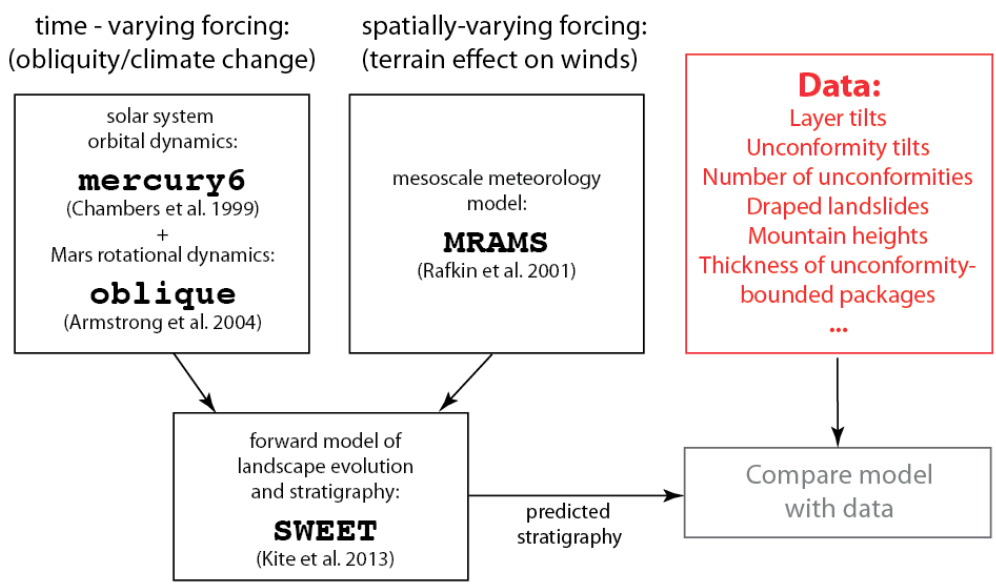

**Fig. A1.** Sketch of `SOURED` model. Combining the MRAMS output with the obliquity forcing, we use the stratigraphic forward model to predict the structure of the mounds.

## A.2. Stratigraphic forward model.

The purpose of our forward stratigraphic model is generate basin stratigraphies for comparison with observations. Earth models with the same purpose (but different physics) include `SedSim` (Griffiths et al. 2001) and `Dionisos` (Csato et al. 2014). Our forward stratigraphic model is 2D (one horizontal dimension and one vertical dimension), with a nominal resolution of ~1 km in the horizontal dimension and 1 Myr in time. Our model does not attempt to resolve processes operating at shorter scales of space and/or time.

The model is modified after the Slope-Wind Enhanced Erosion and Transport (`SWEET`) model of Kite et al. (2013a), with significant enhancements to incorporate parameterized erosion estimators obtained from mesoscale models and time-varying climate forcing (Fig. A1). In `SWEET`,

$$dz/dt = D - e_E \qquad (1)$$

where $D$ is deposition rate and $e_E$ is erosion rate,

$$e_E = k_E U^\beta \qquad (2)$$

where $k_E$ is an erodibility parameter, $U$ is wind shear-stress, and $\beta$ is in the range 2-4 for wind-erosion processes (Kok et al. 2012). The threshold for sediment mobilization is omitted, which is a large simplification. Since the gap between the fluid threshold for saltation initiation and the impact threshold for saltation cessation is so large on Mars, the threshold is very uncertain. Large values of $\beta$ produce a similar pattern of normalized wind erosion to large values of the mobilization threshold. Therefore, combining the threshold with $\beta$ is a reasonable simplification. In earlier work (Kite et al. 2013a) we treated the relative importance of slope winds ($U_s$) and background or "synoptic" winds $U_0$ as a free parameter,

$$U = U_0 + U_s \qquad (3)$$

where $U_0$ could be varied. Here we remove the free parameter $U_0$ by calculating erosion estimators directly from cell-by-cell mesoscale model output,

$$e_E(\beta, s) = k_E(\beta) \frac{1}{N_s} \frac{\Delta t}{(t - t_{su})} \sum^{s=S} \sum_{t_{su}}^{t} \tau_{s,t}^{\beta} \qquad (4)$$

where $N_s$ is the number of grid cells with slopes in the range of interest, $t$ is total elapsed time, $t_{su}$ is spin-up time ($t - t_{su}$ is always an integer number of sols), $\Delta t$ is timestep, $\tau_s$ is the instantaneous surface shear stress (in Pa), and $\beta$ is from equation (2). In practice we use a log-linear fit to the cell-by-cell data to get a smooth relationship between slope and erosion (Appendix B). $k_E$ is adjusted to match the height of observed mounds. In the limit where erosion depends only on local slope (modeled here), and where $e_E \sim 0$ for $s = 0$, the model will tend to produce a cone (or triangular prism) of sedimentary rocks whose side-slope is $dz/dx = (D/k_E)^{(1/\beta)}$.

`SWEET` does not conserve mass locally. Instead, material is added from distant sources (e.g. by airfall), and eroded material is removed to a distant sink (e.g. the Martian lowlands; Grotzinger & Milliken 2012). Layers in the model are assumed to be indurated (mobile sand is assumed to be topographically superficial or to have a geologically short residence time). Because induration probably involves cementation by mineral precipitation from aqueous fluids, long-term secular decline in Mars' ability to form sedimentary rocks (due to, for example, water loss and $CO_2$ loss) means that the model is most applicable to Early Mars.

### A.3.  Orbital dynamics model and obliquity model.

The purpose of our orbital dynamics model and obliquity model is to generate an ensemble of realistic 3.1-Gyr-long obliquity tracks for Mars (Kite et al. 2015). We generated $\varphi$ tracks using `mercury6` (the N-body code of Chambers 1999), and the obliquity code of Armstrong et al. (2004, 2014). For each >3 Gyr long 8-planet solar system integration ($n = 37$) (the combined eccentricity pdf from these integrations is very similar to that of Laskar et al. 2004), we seeded 24 Mars $\varphi$-tracks drawing the initial $\varphi$ from the long term distribution of Laskar et al. 2004. From the ensemble, we selected those $\varphi$-tracks which ended (after 3.1 Gyr) in the range 20°-35° (consistent with present-day Mars $\varphi$). The figures in this paper show a subset of the stratigraphic output forced by those $\varphi$-tracks, chosen to illustrate a range of common stratigraphic outcomes.

## Appendix B. Mesoscale model.

### B.1. Mesoscale model input.

The purpose of our mesoscale modeling work is to verify that wind stress increases with topographic slope. We also seek the 'slope enhancement factor' – to what extent is erosion rate (assumed to scale as wind stress to some power $\beta$) faster on steep slopes than on flat slopes within craters/canyons? We have already verified (Kite et al. 2013a), using the MarsWRF model (Toigo et al. 2012, Richardson et al. 2007), that the strongest winds are on the steepest slopes for a simulation of one year's winds at Gale crater. Here we use the Mars Regional Atmospheric Modeling System (Rafkin et al. 2001) to extend our earlier results through exploring a range of idealized topographies (Tyler & Barnes 2015, Day et al. 2016). MRAMS is derived from the terrestrial RAMS model (Mahre & Pielke 1976). MRAMS has been used to model the entry and descent of all NASA Mars landers subsequent to Pathfinder (Michaels & Rafkin 2008). We use a horizontal resolution of 4.4 km, and a vertical resolution varying from 15 m near the surface to >1 km at high altitude. A realistic diurnal cycle in insolation is imposed (including planetary-scale thermal tides). Our runs are carried out at 6 mbar; the pattern of wind forcing should be similar for other atmospheres that are thin enough for a large day-night cycle in surface temperature. The orbital parameters are for modern Mars, but the diurnally-reversing mesoscale circulation should operate similarly at high obliquity (the background winds may be stronger; Haberle et al. 2003, Newman et al. 2005). Boundary conditions are supplied by the NASA Ames Mars General Circulation Model (MGCM) (Haberle et al. 1993). For this project, we modify MRAMS to simulate idealized craters and idealized canyons. We use smoothed background topography, and insert oblong canyons of width ~130 km and length ~350 km and depth ~4.5 km. We run the model both without mounds, and for canyons containing mounds of 100% of the full height of the canyon (Fig. B1). These runs correspond to idealized topography for a large canyon hosting a large mound (e.g. Candor, Hebes, Ophir). Separately, we insert 4.5 km-deep, 155 km-diameter axisymmetric craters (Fig. B1), with their corresponding mounds. These runs correspond to idealized topography for a large crater hosting a large mound (e.g. Gale crater, Nicholson crater). We ran for 5.7 day-night cycles for solar longitude $L_s$= {30°, 90°, 150°, 180, 210°, 270°, 330°}. The first 1.7 sols are discarded as spinup. Simulated crater/canyon latitude is ~5°S.

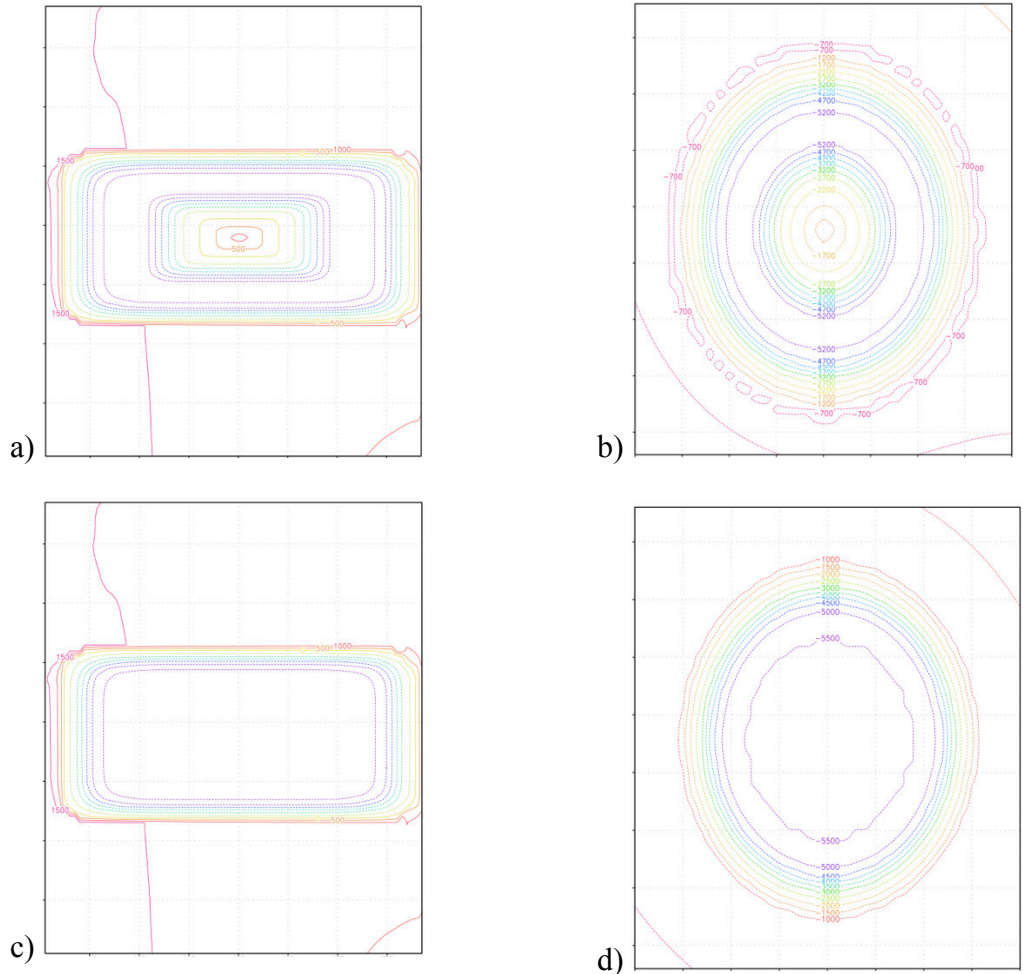

**Fig. B1.** Topographies investigated using MRAMS simulations. Contours at 500m intervals. a) Rectangular canyon, width ~130 km and length ~350 km, with full-height mound. b) Crater, 155 km diameter, with full-height mound. c) Flat-floored canyon without sediment infill. d). Flat-floored crater without sediment infill.

## B.2. Mesoscale model output.

Our MRAMS runs confirm that the strongest winds within craters/canyons are associated with diurnally-reversing (anabatic/katabatic) flows, and are located on the steepest slopes (Fig. B2). Terrain-controlled circulation dominates the overall circulation inside our idealized craters and canyons (consistent with Tyler & Barnes 2015) (Fig. B3). The importance of slope winds in our idealized-topography runs is somewhat offset for real craters and canyons by regional effects (e.g. the planetary topographic dichotomy boundary, Rafkin et al. 2016). To simplify the analysis, we assume cell-scale (4km-scale) control of terrain on wind stress. Gridcells inside a canyon or crater are generally *less* windy than on the plateau surrounding the depression (Fig. B3). This is partly because the plateau is subject to the morning "surge" of air moving away from the canyon (Tyler & Barnes 2015). However, *within* the crater/canyon, wind stress is about 5× greater for 15° slopes than for flat surfaces. Points just below the rim of the crater/canyon have stronger wind stress than expected for their slope, because they participate in the morning "surge" of air moving away from the crater/canyon. The scatter of mean wind stress is about a factor of 2. We use the crater output; the same trends were found for canyons as for craters.

How we get from mesoscale model output to erosion estimators: Even if our wind models perfectly represented wind stresses inside Mars craters/canyons, that would not be enough to correctly diagnose the rate of aeolian erosion of bedrock. Aeolian erosion of rock is a multi-step process (Shao et al. 2008, Kok et al. 2012), and it is difficult to determine the rate-limiting step from orbit. Possibilities include breakdown of sedimentary layers to wind-transportable fragments by weathering and/or volume changes associated with hydration state changes (e.g., Chipera & Vaniman 2007); physical degradation by mass wasting, combined with aeolian removal of talus (Kok et al. 2012); aeolian erosion of weakly salt-cemented sediments (Shao, 2008); and aeolian abrasion of bedrock (Wang et al., 2011). Rather than attempting to directly predict erosion rate, we use the strong evidence for geologically-recent wind erosion of the mounds (e.g. Day et al. 2016) to establish the feasibility of aeolian sculpting of the mounds, and we use the wind models to get the pattern of past wind erosion. This requires us to accept two limitations:

1. The relationship between wind stress and erosion rate will vary depending on both past atmospheric pressure and the details of the erosion process. We parameterize this uncertainty by a power-law exponent, $\beta$.
2. Wind erosion is not carried out by the wind directly, but by sand grains carried by the wind (at least for most erosion processes), and sand grains are not tracked by the model This is directly analogous to the tools-and-cover problem in modeling the evolution of Earth's mountains (section 7).

To get the relationship between wind stress and slope, we tried fitting the unbinned data with various functions (exponential, two-exponential, power law, polynomial, e.t.c). The most visually satisfying fit is a log-linear function. The fit suffices to capture the basic tendency for wind erosion within craters and canyons to be stronger on steep slopes than on gentle slopes, by a factor of between ~4 (if erosion is proportional to mean wind stress) and >10 (if erosion is proportional to wind stress raised to the fourth power).

We defined erosion estimators from the MRAMS model output on a per-gridcell basis as follows

$$e_E(\beta) = \frac{k_E(\beta)}{(t - t_{su})} \int_{t_{su}}^{t} \tau_{sw}^{\beta} \, dt$$
(5a)

$$e_E(\beta, s) = k_E(\beta) \, 10^{k_1(\beta)s + k_2(\beta)}$$
(5b)

We obtained $e_E$ by regression using a log-linear fit where $e_E = 10^{(k_1 s + k_2)}$, where $s$ is slope. We did this for the 100%-mound simulation (mound-in-crater) (Fig. B3). The erosion estimators are (for $\beta =1$, i.e. erosion proportional to mean wind speed) $k_1 = 3.08(2.84,3.31)$, $k_2 = -3.37(-3.41,-3.33)$, (for $\beta =2$) $k_1 = 5.34(4.87,5.82)$, $k_2 = -6.34(-6.42,-6.26)$, (for $\beta =3$) $k_1 = 7.44(6.69,818)$, $k_2 = -9.13(-9.26,-9.01)$ and (for $\beta =4$) $k_1=9.42(8.39,10.44)$, $k_2 = -11.81(-11.98,-11.64)$. Here, the brackets give the formal confidence interval of the fits. The choice of erosion estimator depends on the paleoatmospheric pressure and on the mechanism of erosion. For low atmospheric pressure, $u*_{cr}$ (the surface-stress threshold for sand motion) approaches the maximum wind speed, and $\beta \rightarrow \infty$ (i.e. erosion only responds to the very strongest gusts). Sand dunes on Mars today are in active motion (Bridges et al. 2012a), so $\beta < \infty$. When atmospheric pressure was higher earlier in Mars history, $u*_{cr}$ would become small compared to frequently-encountered wind speeds. Under those circumstances, $2 < \beta < 4$ is appropriate. We used $\beta = 3$ to make the plots shown in this paper. We carried out sensitivity tests changing the $\beta$ parameter, finding no qualitative difference for $2 < \beta < 4$ (after adjustment for each $\beta$ of the dimensionless deposition rate in order to match observed mound heights).

The past terrain-averaged erosion rate is effectively a free parameter in our model. We find good results with maximum past rates that are comparable to Earth wind erosion rates, that agree with previous calculations of peak present-day Mars wind erosion rates (Bridges et al. 2012b), and that are 1 order of magnitude greater than typical present-day Mars sedimentary rock wind erosion rates (Golombek et al. 2014, Kite & Mayer submitted), consistent with higher atmospheric pressure (or weaker rocks) in the past.

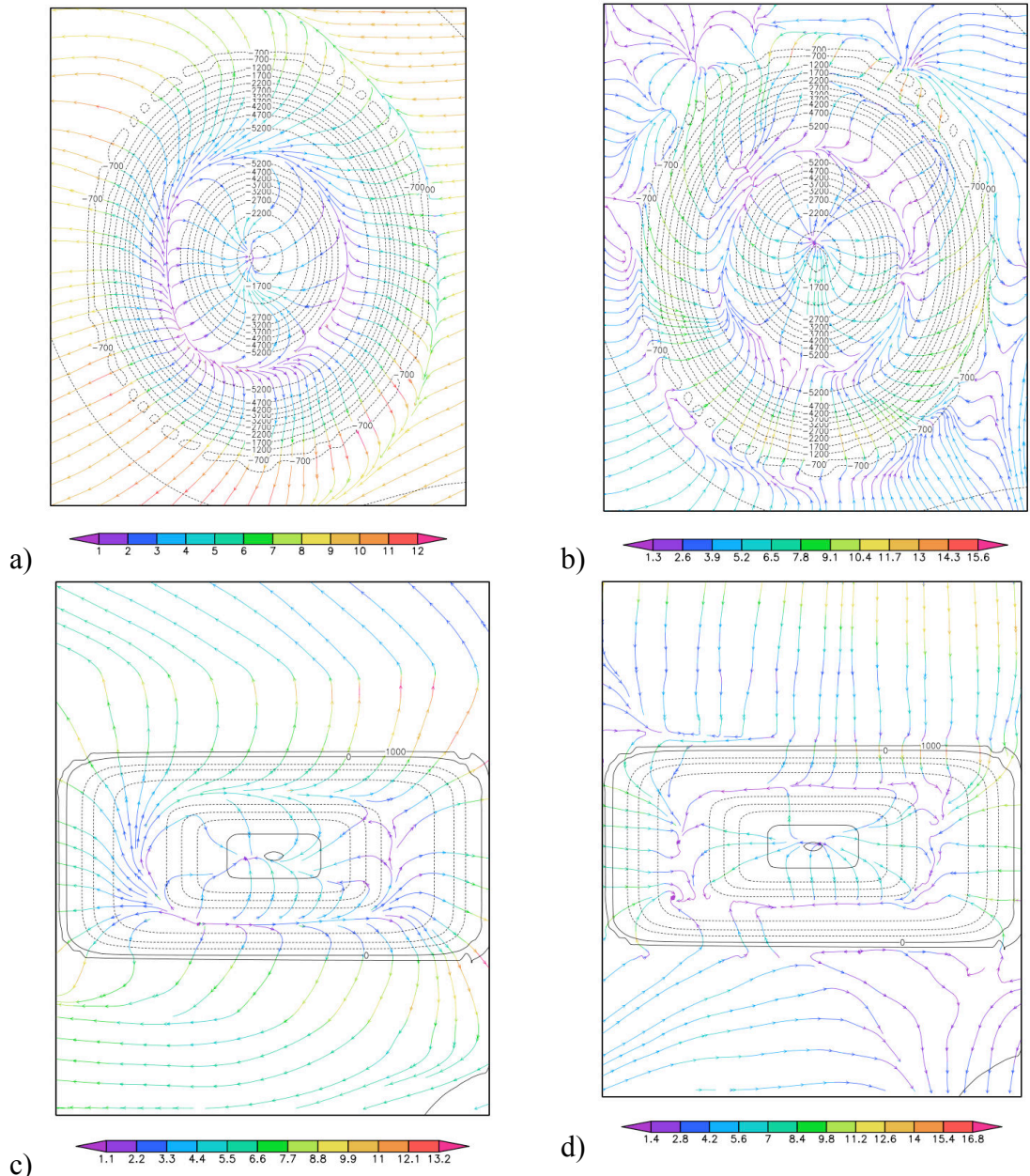

**Fig. B2.** Winds are strong on crater/canyon walls and on mound flanks, but weak in the moat. (a) Snapshot of daytime flow in idealized-topography crater, dominated by upslope (anabatic) winds. Contours (m) are elevation. Colors are wind speed (m/s) at 15m elevation. Topographic contour interval 500m. (b) Snapshot of nighttime flow in idealized-topography crater, dominated by downslope (katabatic) winds. Topographic contour interval 500m. (c) Snapshot of daytime flow in idealized-topography canyon, dominated by upslope (anabatic) winds. Topographic contour interval 1000m. (d) Snapshot of nighttime flow in idealized-topography canyon, dominated by downslope (katabatic) winds. Topographic contour interval 1000m. (This figure was produced using the Grid Analysis and Display System, GrADS: http://cola.gmu.edu/grads/).

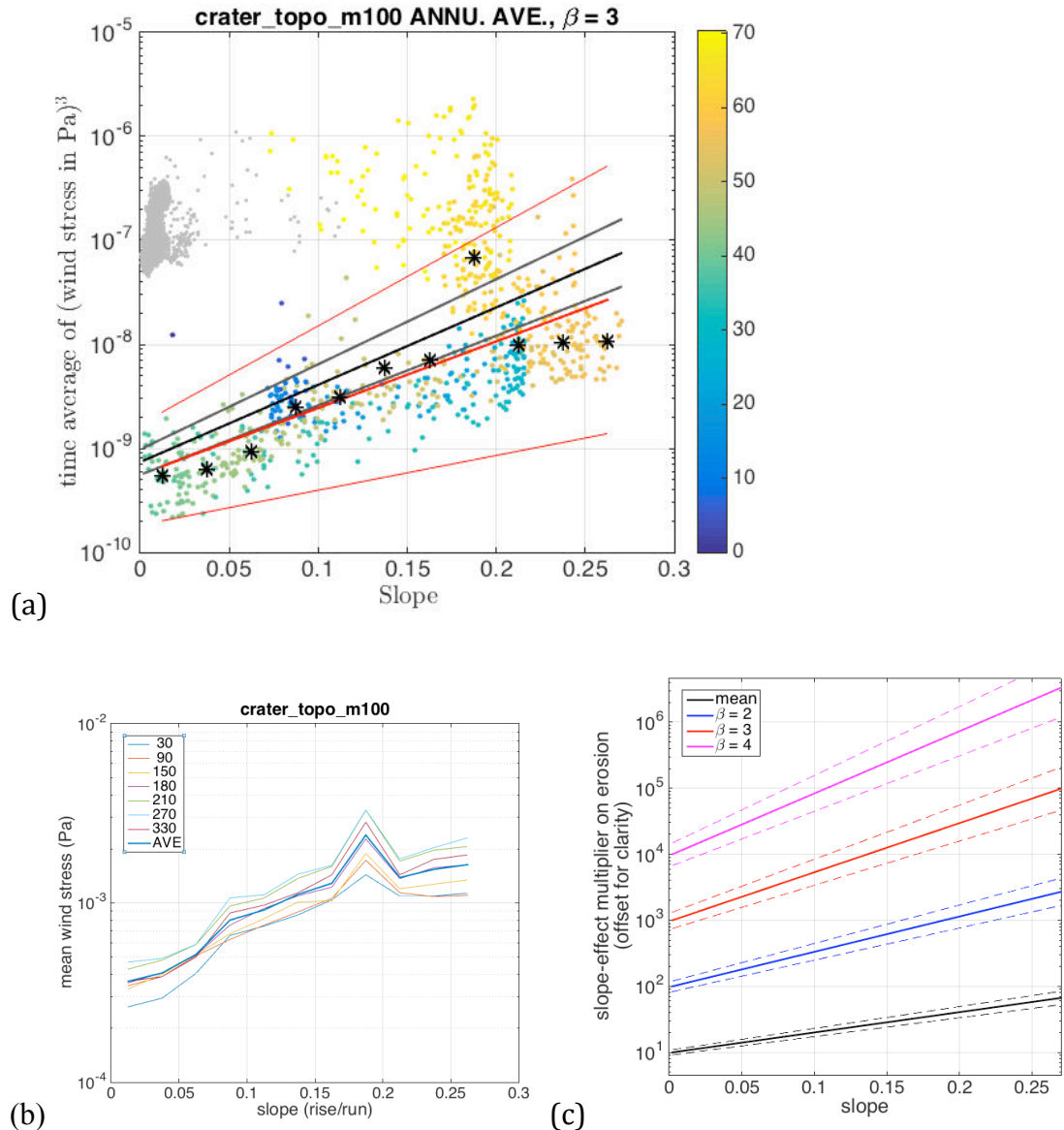

(a)

(b)                                    (c)

**Fig. B3.** MRAMS run-integrated output. (a) Maximum wind stress from the last 4 sols of model runs, for topographic boundary conditions featuring large central mounds. Model output is binned according to grid cell slope, and the median for each bin is shown (stars). Colors correspond to distance from mound center in kilometers. The black line (and gray error bars) corresponds to the best log-linear fit to all data. The red line (and red error bars) correspond to the best log-linear fit to the binned data (asterisk symbols). (b) Mean wind stress from the last 4 sols of model runs at different seasons, for topographic boundary conditions corresponding to a full-height central mound inside a crater. The numbers in the legend correspond to the $L_s$ (Martian season) of each run. (c) The overall best fits to the slope-effect multiplier (offset for clarity; for slope = 0, the slope-multiplier effect is 1 by definition).

# References.